\documentclass{jfm}
\usepackage{graphicx}
\usepackage{epstopdf, epsfig}
\usepackage{amsmath}

\usepackage{lineno}

\usepackage{float}

\usepackage{xspace}

\makeatletter
\DeclareRobustCommand\onedot{\futurelet\@let@token\@onedot}
\def\@onedot{\ifx\@let@token.\else.\null\fi\xspace}

\def\eg{\emph{e.g}\onedot} 
\def\ie{\emph{i.e}\onedot}

\def\etal{\emph{et al}\onedot}
\makeatother

\shorttitle{Similarity of wake characteristics for yawed wind turbines of utility-scale}
\shortauthor{Zhaobin Li and Xiaolei Yang}

\title{Similarity of wake characteristics for yawed wind turbines of utility-scale}

\author{Zhaobin Li\aff{1,2},
 \and Xiaolei Yang\aff{1,2}\corresp{\email{xyang@imech.ac.cn}}}

\affiliation{\aff{1}The State Key Laboratory of Nonlinear Mechanics, Institute of Mechanics, Chinese Academy of Sciences, Beijing 100190, China
\aff{2}School of Engineering Sciences, University of Chinese Academy of Sciences, Beijing 100049, China}

\begin{document}

\maketitle

\begin{abstract}
This work is dedicated to studying the influence of yaw angle on the wake characteristics of a 2.5 MW utility-scale wind turbine using large-eddy simulations with the turbine's blades and nacelle parameterized as actuator surfaces. Four different yaw angles are simulated, \ie{}, $\gamma=0^{\circ},10^{\circ},20^{\circ},30^{\circ}$ for three different tip-speed ratios $\lambda = 7,8,9$. Similarities  are observed for both  time-averaged velocity fields and the turbulence statistics of turbine wakes.  Specifically, it is found that the temporal average of the wake deflection, the velocity deficit  and the transverse velocity, the Reynolds stresses, and the wake width defined using the velocity deficit can be properly scaled using the characteristic length and velocity scales depending on the thrust on the rotor and yaw angles. The width of the region influenced by the transverse velocity, the PDF of the instantaneous wake position, and the standard deviation of wake position and wake width, on the other hand, are scaled well using the rotor diameter independent of the yaw angles and tip-speed ratios. These results suggest that the wake of a yawed wind turbine can be decomposed into the streamwise and transverse components, and the turbine-added turbulence and the meandering motion of the wake can be decoupled from the wake deflection for the considered cases.

\end{abstract}


\section{Introduction}\label{sec:introduction}

Wind turbine wakes featured by low wind speed and high turbulence intensity, reducing the power outputs and increasing the fatigue loads of downwind turbines, can significantly impact the overall performance of wind farms \citep{barthelmie2009WakeMeasuring,thomsen1999fatigue}.  To mitigate the negative impacts of turbine wakes, advanced turbine control strategies have been developed in the literature, \eg{}, axial induction factor control \citep{annoni_analysis_2016},  yaw-based wake control \citep{wim_munters_dynamic_2018,hoyt_wind_2020} and individual blade pitch control \citep{ossmann_load_2017}. Understanding how these control strategies affect the characteristics of turbine wakes is critical for their implementation in utility-scale wind farms. In this work, we focus on the wake flow characteristics of yawed wind turbines, in which the wake is redirected by deliberately creating a misalignment between the rotor’s axis and the inflow directions \citep{gebraad_maximization_2017,Kragh_yaw_load}. Specifically, we investigate the similarity of the time-averaged velocity, turbine-added Reynolds shear stress and meandering of the wake of a utility-scale wind turbine four a series of yaw angles and tip-speed ratios (TSR). 

The wake deflection is one key feature of the wake for a yawed turbine. It has been extensively studied in the literature in terms of its origin and characteristics. Using hot-wire measurement in a wind tunnel of a small wind turbine of rotor diameter $D = 0.12$ m, \cite{medici2006measurements} found that the wake deflection is caused by the transverse velocity, which convects the wake in the transverse direction. Later, more details on the transverse velocity were revealed by \cite{howland2016wake} using simulations of a yawed actuator disk. Their simulation results showed that the transverse velocity is non-uniformly distributed on the cross-section and can be described by a counter-rotating vortex pair (CVP) with vortex centers located above and below the wake center, which not only convects the wake laterally but also deforms it into a kidney-shape.  \cite{bastankhah_experimental_2016} explained the formation of the CVP based on the continuity equation. Moreover, from the wind tunnel experiment of a wind turbine of $D = 0.15$ m, they showed the interaction between the CVP and the hub vortex in the wake, which adds an extra top-down asymmetry to the wake cross-section.  
\cite{bartl2018wind} conducted wind tunnel measurements of a yawed-turbine model ($D = 0.90$ m) for different inflow turbulence and found that the kidney-shape deformation of the wake cross-section is much alleviated when increasing the inflow turbulence intensity from 2\% to 10\%. 

To facilitate the design and operation of wind farms, analytical wake models have been derived in the literature for fast prediction of wake center-line deflections.  In the model of \cite{jimenez_yaw_2010}, the far wake's transverse velocity for computing wake deflections is related to the lateral component of the thrust on rotor based on the momentum theory. Later, it was found that the model by Jim\'{e}nez \etal{} may over-predict the wake deflection \citep{jimenez_yaw_2010,shapiro_modelling_2018}. To solve this problem,  \cite{shapiro_modelling_2018} proposed a model based on the analogy of a yawed wind turbine to a finite elliptical wing. In the proposed model, the transverse velocity in the wake is assumed to be equal to the the downwash on the wing created by the CVP according to Prantl's lifting line theory, predicts the wake deflection 1/2 as large as that of \cite{jimenez_yaw_2010}. 
Validation of Shapiro \etal{}'s model was carried out using the simulation results of an actuator disk model and the wind tunnel measurement of \cite{bastankhah_experimental_2016}.  It was found that the key to successfully predicting the wake deflection is to accurately compute the transverse velocity, which gradually decreases via the downstream distance to the turbine. Both Jim\'{e}nez \etal{}'s model and Shapiro \etal{}'s  model attribute the gradual decrease of lateral velocity to the wake expansion, and assume that the wake widths defined by the streamwise velocity and the transverse velocity are equal, so that the wake width computed from the streamwise velocity can be employed for the transverse  velocity. The above assumption is also adopted in other analytical wake models \citep{bastankhah_experimental_2016,qian_new_2018}. However, it has not yet been proved and may introduce error as pointed out by \cite{jimenez_yaw_2010}. To this end, a part of this work will be dedicated to verifying if the wake widths defined by the velocity in the streamwise and spanwise directions are equal. 

With the aforementioned models, the wake deflection can be predicted with reasonable accuracy. However, to realize yaw-based wake steering methods for optimal wind farm design and operation, fast analytical models need to be developed for predicting more features of the wake of a yawed wind turbine, which includes both time-averaged quantities (\eg{}, wake width and time-averaged  velocity deficit), and instantaneous quantities (\eg{}, instantaneous wake influencing region). Most existing analytical models for wakes of yawed turbines have been focused on the deflection and velocity deficit of the time-averaged wake. For instance, the analytical wake model of Bastankhah \& Porte-Agel (2016), which was developed based on the self-similarity of the streamwise velocity and the skew angle observed in the far wake, can predict the velocity distribution and wake deflection of yawed wind turbines.  On the other hand, high-fidelity models, \eg{}, large-eddy simulation (LES) with actuator surface/line models for turbine blades, can accurately predict various aspects of turbine wakes. However, it is still not feasible to use high-fidelity models for the optimization of wind farm design and operations due to the expensive simulation cost and the large number of cases to be simulated in an optimization process. On the other hand, similarity of turbine wakes has been observed from the LES results for different wind turbine designs \citep{foti2018similarity}, different inflow \citep{yang_3Dhill_2015}, and different turbine operational conditions \citep{yang_meanderingLES_2019}. If the similarity also exists in the turbine wakes at different yaw angles, then there is a potential that fast models can be derived for predicting various aspects of wakes of yawed turbines based on such similarity and several typical high-fidelity simulations.



As a step towards the development of a general wake model for yawed wind turbines, the objective of this work is to examine the similarity of turbine wakes for different yaw angles.  Specifically, we conduct a series of simulations of the EOLOS wind turbine \citep{hong2014natural,chamorro2015turbulence} at four different yaw angles ( \ie{} $\gamma = 0^o, 10^o, 20^o, 30^o$) for different tip-speed ratios  under fully developed turbulent inflow. The wind turbine wake is simulated using LES with the wind turbine's blades and nacelle modeled as actuator surfaces \citep{yang2018ASMethod}. The time-averaged wake velocity field, turbine-added Reynolds stress, and the instantaneous wake positions from the simulations are systematically examined and different scaling factors are derived to describe the similarity of these wake characteristics. 


The rest of the  paper is structured as follows. In section 2, we describe the employed numerical methods and the simulation setup. Then in sections 3 and 4, we present results on the similarity of time-averaged wake velocity field and statistics of the wake turbulence, respectively. At last, we summarize the findings from this work and draw conclusions in section 5. 
 
\section{Numerical methods and simulation setup}\label{sec:numerical_method}

\subsection{Flow solver}
The turbulent flow is solved using the LES module of the Virtual Flow Simulator code (VFS-Wind,  \cite{yang2015largeEddy}), in which the governing equations are the filtered incompressible Navier-Stokes equations in curvilinear coordinates shown as follows:

\begin{align}
	J \frac{ \partial U^i }{ \partial \xi^i } & =0, \\
	\frac{1}{J}\frac{\partial U^j}{\partial t} & = \frac{\xi^i_l}{J}\left( -\frac{\partial }{\partial \xi^j} \left(U^ju_l\right)+\frac{\mu}{\rho}\frac{ \partial }{\partial \xi^j}\left(\frac{g^{jk}}{J}\frac{\partial u_l}{\partial \xi^k}\right) -\frac{1}{\rho}\frac{\partial}{\partial \xi^j}\left( \frac{\xi^j_l p}{J}\right)-\frac{1}{\rho}\frac{\partial \tau_{lj}}{\partial \xi^j} +f_l\right), \label{eqn:ns}
\end{align}
where $i,j,k,l=\{1,2,3\}$ are the tensor indices,  $\xi^i$ is the curvilinear coordinates related to the Cartesian coordinates  $x_l$ by  the transformation metrics $\xi^i_l  = \partial \xi^i/\partial x_l $.  $J$  denotes the Jacobian of the geometric transformation, $U^i = \left(\xi^i_l / J\right)u_l$ is the contravariant volume flux with $u_l$ the velocity in Cartesian coordinates, $\mu$ denotes the  dynamic viscosity, $\rho$ is the fluid density, $g^{jk} = \xi^j_l\xi^k_l$ are the components of the contravariant metric tensor, and $p$ is the pressure, and $f_l$ are body forces introduced  by the actuator type wind turbine model. In the momentum equation, $\tau_{ij}$ is the subgrid-scale (SGS) stress modeled following \cite{Smagorinsky} as follows,
\begin{equation}
	\tau_{ij}-\frac{1}{3}\tau_{kk}\delta_{ij}=-\mu_t \overline{S_{ij}},
	\end{equation}
where $\overline{S_{ij}}$ is the filtered strain-rate tensor with $\overline{(\cdot)}$ denoting the grid filtering operation and $\mu_t$ is the eddy viscosity computed by 
	\begin{equation}
	\mu_t = C_s \Delta^2 |\overline{S}|,
	\end{equation}
	where $\Delta$ is the filter width, $ |\overline{S}| = (2 \overline{S_{ij}} \overline{S_{ij}})^{1/2} $  is the magnitude of the strain-rate tensor and $C_s$ is the Smagorinsky constant computed via the dynamic procedure of \cite{germano1991subGridModel}.

The governing equations are discretized on a structured curvilinear grid. A second-order accurate central differencing scheme is used for space discretization. A second-order fractional step method \citep{ge2007numerical} is employed for temporal integration. The momentum equation is solved with a matrix-free Newton-Krylov method \citep{knoll2004jacobian}. The pressure Poisson equation for satisfying the continuity equation constraint, is solved using the Generalized Minimal Residual (GMRES) method with an algebraic multi-grid acceleration \citep{saad1993flexible}.

\subsection{Wind turbine parameterization method}

As the length scale of the wind turbine's wake ($\propto D$ m, where $D$ is the rotor diameter ) is often more than two orders of  magnitude larger than the thickness of the boundary layer on the blade ($\approx 1$ cm  for a turbine of $D \approx 100$ m  operating in region II~\citep{yang2018ASMethod}), it is extremely expensive to simulate the blade aerodynamics by directly solving Navier-Stokes equations in wake simulations, such that the aerodynamics of wind turbine are often parameterized using actuator disk \citep{chattot2014actuator}, actuator line \citep{sorensen2002AL} or actuator surface \citep{shen2009ASM,yang2018ASMethod} models. In this work, a class of well validated actuator surface (AS) models for turbine blades and nacelle  proposed by \cite{yang2018ASMethod} is employed . The AS model represents each rotor blade with a simplified two dimensional surface defined by the chord length and the twist angle at different radial locations. The lift and drag forces $\mathbf{L}$ and $\mathbf{D}$ at each radial location are determined using the tabulated airfoil data using the local instantaneous relative incoming velocity as follows:

\begin{equation}
\mathbf{L} = \frac{1}{2}\rho C_\text{L}c|V_\text{ref}|^2 \mathbf{e_\text{L}} 
\end{equation}
and 
\begin{equation}
\mathbf{D} = \frac{1}{2}\rho C_\text{D}c|V_\text{ref}|^2 \mathbf{e_\text{D}}, 
\end{equation}
where  c is the chord length, $V_\text{ref}$ is the instantaneous incoming velocity relative to the rotating blade at the computing point, $\textbf{e}_\text{L}$  and $\textbf{e}_\text{D}$ are unit directional vectors of lift and drag forces, and $C_\text{L}$ and $C_\text{D}$ are the lift and the drag coefficients defined in 2D airfoil tables as a function of Reynolds number and the angle of attack. Corrections including the 3D stall delay model \citep{du19983Dstall} and the tip loss correction \citep{shen2005tiploss2} are applied.
With the  computed \textbf{L} and \textbf{D}, the body force in \eqref{eqn:ns} is calculated by uniformly distributing the forces along the chord as follows:
\begin{equation}
\mathbf{f} = (\mathbf{L}+\mathbf{D})/c.  
\end{equation}
A smoothed discrete delta function \citep{yang2009Kernel} is employed for transferring quantities between the actuator surface and the background grid for solving the flow. 

\subsection{Simulation Setup}

The numerical experiment consists of three sets of wind turbine wake simulations with three different TSRs ($\lambda \in \{7,8,9\}$). For each TSR, the yaw angle $\gamma$, which is defined as the misalignment between the inflow and the rotor axis (shown in figure \ref{fig:computationalDomain}), varies among $\{0^o, 10^o, 20^o, 30^o\}$. One extra case without wind turbine is also simulated to provide the reference turbulent boundary layer flow for computing quantities such as the turbine-added turbulence kinetic energy and instantaneous velocity deficit for computing the instantaneous wake positions.  All the simulations are conducted with the same setup, including the same boundary, initial conditions, and time step to produce synchronized results enabling the comparison of instantaneous flow fields. The physical time simulated is approximately 75 minutes to achieve statistical convergence. Details on the employed wind turbine, computational domain and boundary conditions, and the turbulent inflow generation  are provided in the following subsections. 

\subsubsection{Wind turbine and operation condition}

We simulate the three-blade horizontal-axis Clipper Liberty 2.5 MW  wind turbine located at the EOLOS wind energy research field station at the University of Minnesota, USA. The rotor diameter is $D=96$ m, the hub is at $z_\text{hub} = 80$ m, and the nacelle has a cuboid-like shape with its dimensions being approximately equal to $5.3 \;\text{m} \times 4.7 \;\text{m} \times 5.5 \;\text{m} $. The tower is conical with diameters varying from 3.0 m at the top to 4.1 m at the bottom. More information about the EOLOS wind turbine can be found in \cite{hong2014natural} and \cite{chamorro2015turbulence}. In the simulated cases, the incoming wind speed at hub height is $U_{\infty} = 9 $ m/s with the corresponding standard deviation of the streamwise velocity fluctuations $\sigma_u/U_{\infty} = 0.08$. The Reynolds number based on $D$ and $U_{\infty}$  is $5.7 \times 10^{7}$. In each case, the rotor rotates at a fixed TSR. The TSR for yawed wind turbines is defined with respect to the flow velocity normal to the rotor sweeping plane, as follows: 
\begin{equation}
\lambda  = \frac{\Omega R}{U_d} \label{eqn:TSR}, 
\end{equation} 
with $\Omega$ the angular velocity of the rotor, $R = 48$ m is the radius of the rotor, and  
\begin{equation}
{U_d} = U_\infty \cos \gamma    
\end{equation}
is the inflow velocity projected in the rotor's axis direction. This projected velocity is employed because a yawed wind turbine achieves the optimal energy conversion, controlled by TSR with respect to $U_d$ as shown by \cite{burton2011wind}. 

\subsubsection{Computational domain and boundary conditions}

\begin{figure}
	\centering
	\includegraphics[width=0.8\textwidth]{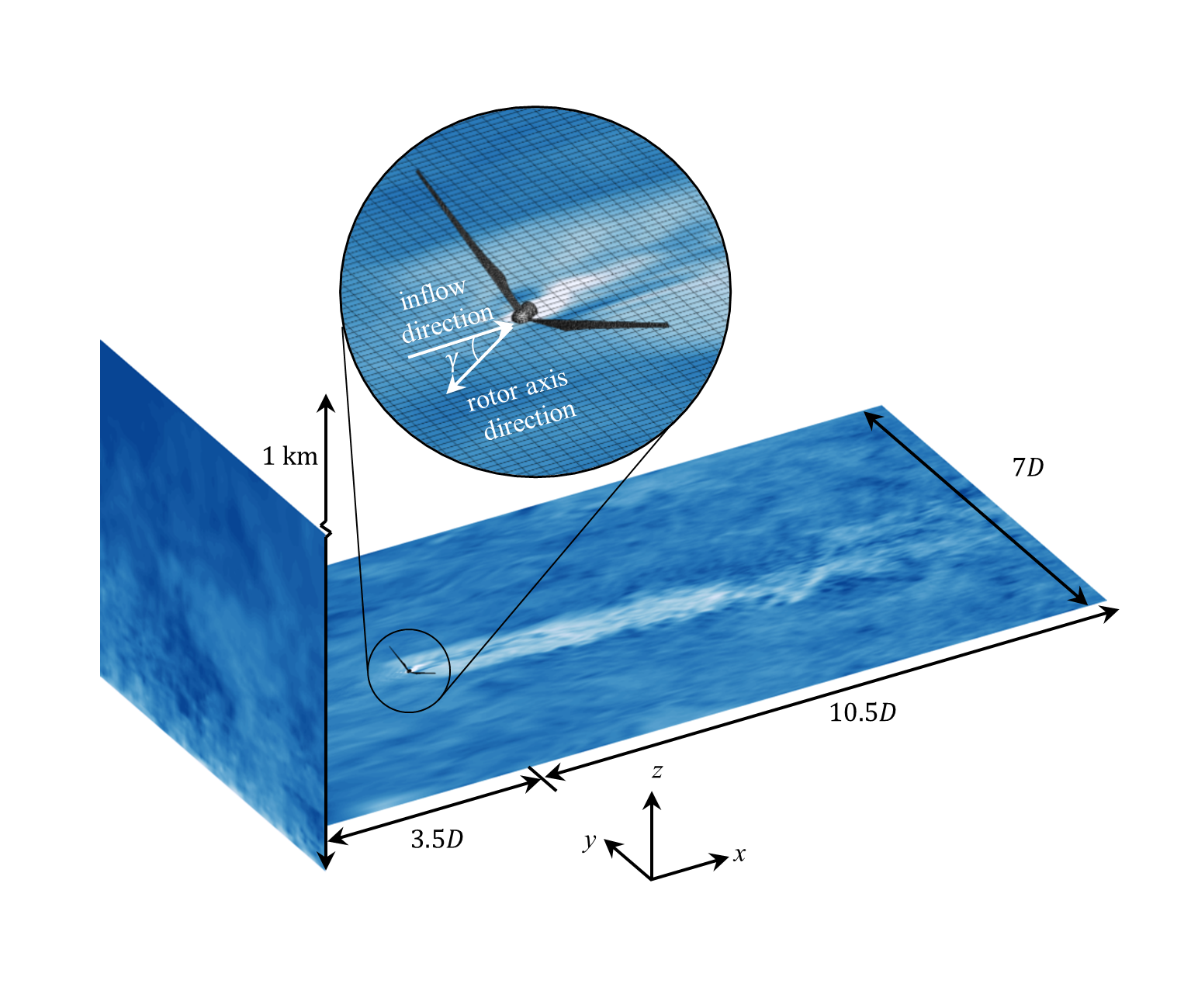}
	\caption{Schematic of the computational setup employed for simulating wake of yawed wind turbine.}
	\label{fig:computationalDomain}
\end{figure}

We employ the same computational domain for all the cases, which is shown in figure \ref{fig:computationalDomain}. The size of the computational domain is $L_x \times L_y \times L_z = 14D \times 7D \times 1~\textrm{km}$, in streamwise ($x$), transverse ($y$), and vertical ($z$) directions, respectively. The origin of the coordinates coincides with the wind turbine's footprint on the ground.  The domain is discretized by a Cartesian grid with number of grid nodes  $N_x \times N_y \times N_z = 281\times 281\times 143 $. The grid  is uniform in the $x,y$ directions with grid spacing $\Delta x = D/20$ and $\Delta y = D/40$. In $z$ direction, the grid is uniform near the ground  ($z \in (0,2D)$) with $\Delta z  = D/40$ and is gradually stretched to the top boundary.  The turbulent inflow generated by a precursory simulation (to be explained in the next section) is prescribed at the inlet ($x = -3.5D$). At the outlet  ($x = 10.5D$), the Neumann boundary condition is applied. A wall model based on the logarithmic law for rough walls is applied on the ground with the roughness length $z_0 = 0.00016$~m, which is typical for calm open sea or snow-covered land at the EOLOS site \citep{hansensurface,gromke2011aerodynamic}. At the lateral and the top boundaries of the computational domain, the free-slip condition is applied.  The size of time step is $ 0.018 D/U_{\infty}$.  The simulation is first run for approximately 40 rotor revolutions to achieve a fully developed state. Then the simulation is continued for another 1000 rotor revolutions (corresponding to physical time 75 minutes) for computing statistics of the wake. The total number of rotor revolutions is computed based on the rotor rotational speed of the $\lambda = 8$ and $\gamma = 0$ case. 

\subsubsection{Turbulent inflow}

The turbulent inflow is generated from a precursory LES with a larger computational domain ($L^{'}_x \times L^{'}_y \times L^{'}_z = 62D \times 46D \times 1~\textrm{km}$) for capturing large scale eddies in the atmospheric boundary layer \citep{wang2016very,liu2019three}. Periodic boundary conditions are applied in horizontal directions.
The top and bottom boundaries of the computational domain are set with the free-slip boundary condition and the logarithmic law  for rough walls, respectively, the same as that in the turbine wake simulations.  In the precursory simulation, the velocity fields on a $y-z$ plane are saved at each time step and then applied at the inlet of the turbine simulations. In case the grid size or time interval in the precursory simulation are different from those of the wind turbine simulations, linear interpolations are carried out. 

\section{Similarity of time-averaged wake characteristics}\label{sec:results_discussion}

\subsection{Variation of the thrust and power via yaw angles}

\begin{figure}
	\centering
	\includegraphics[width=\textwidth]{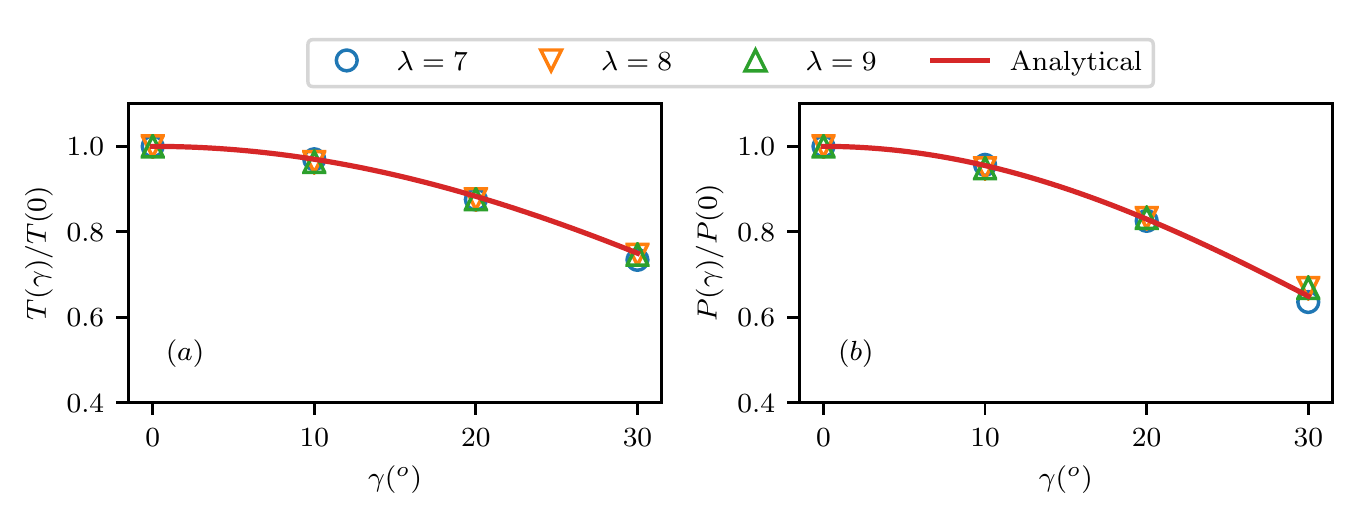}
	\caption{Influence of yaw angle $\gamma$ on the rotor's thrust and power: $(a)$ the normalized thrust $T(\gamma)/T(0)$ (symbols) compared with $\cos^2 \gamma $ (solid line), and $(b)$ the normalized power $P(\gamma)/P(0)$ (symbols) compared with  $\cos^3 \gamma $ (solid line). $T(0)$ and $P(0)$ are the thrust and the power at $\gamma=0$. }
	\label{fig:powerandthrust}
\end{figure}

Before probing into the flow fields in the wake, the variations of the computed thrust (normal to the rotor sweep plane) and power (calculated from the shaft torque) via yaw angles are compared with theoretical predictions in figure \ref{fig:powerandthrust}. To exclude the influence of the TSR on these coefficients, the results are normalized by the value at $\gamma = 0 $ at each corresponding TSR. As seen, all results agree well with the analytical expression obtained using the axial momentum theory \citep{burton2011wind}, which states that the thrust $T$ and power $P$ scale with $\cos^2 (\gamma) $ and $\cos^3 (\gamma)$, respectively. This relation has also been observed in other numerical and experimental studies \citep{krogstad_performance_2012,bartl2018wind}. With this relation, the thrust and power coefficients ($\widetilde{C}_T$ and $ \widetilde{C}_P$), which are defined with respect to $U_d$ and shown as follows,  
\begin{align}
\widetilde{C}_T &=  \frac{T}{\frac{1}{2}\rho A U_d^2} =\frac{T}{\frac{1}{2}\rho A \left(U_\infty\cos\gamma\right)^2}  \label{eqn:thrust},\\ 
\widetilde{C}_P & =   \frac{T}{\frac{1}{2}\rho A U_d^3}= \frac{P}{\frac{1}{2}\rho A \left(U_\infty\cos\gamma\right)^3},
\end{align}  
are independent of the yaw angles and are equal to those in non-yawed cases, because of $T \propto \cos{\gamma}^2$ and $P \propto \cos{\gamma}^3$. The values of $\widetilde{C}_T$ and $\widetilde{C}_P$ for different TSR are shown in table \ref{tab:CtAndCp}. 

\begin{table}
\centering
\begin{tabular}{cccc}
$\lambda$ & 7 & 8 & 9\\
$\widetilde{C}_T$       & 0.654            & 0.710  &  0.711             \\
$\widetilde{C}_P$       & 0.443            & 0.488  &  0.487                 
\end{tabular}
\caption{The modified thrust and power coefficients at different tip-speed ratios. \label{tab:CtAndCp}}
\end{table}

\subsection{Time-averaged velocity fields \label{sect:fieldContour}}

\begin{figure}
	\centering
	\includegraphics{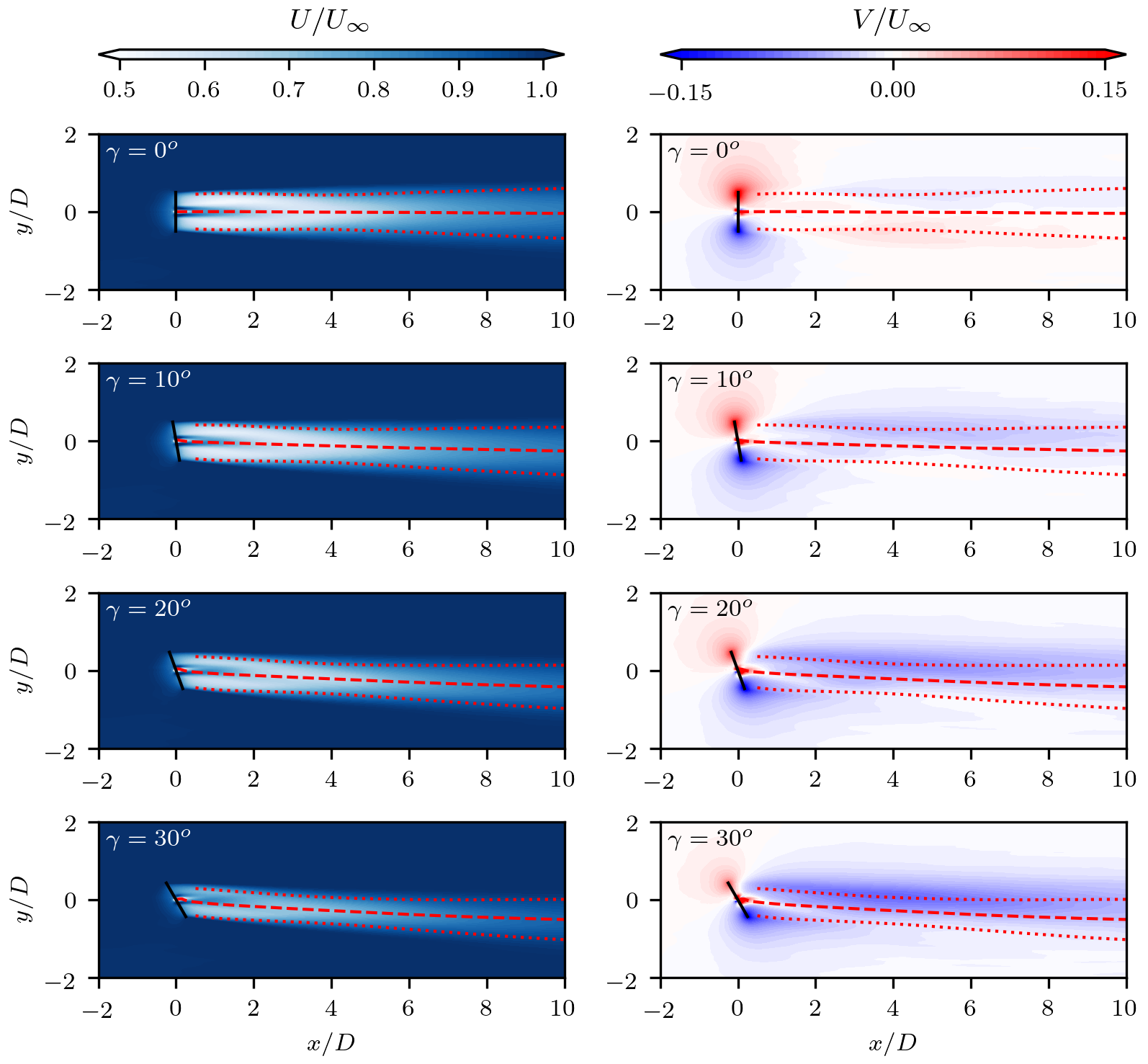}
	\caption{The time-averaged velocity field behind wind turbines (at hub height plane $z=z_\textrm{hub}$). The first column shows the streamwise velocity and the second column shows the transverse velocity. Panels in each column are ordered with increasing yaw angle. The red dashed lines represent the center lines of the streamwise velocity deficit. The red dots denote the wake width where the streamwise velocity deficit is half of that on the centerline. The black solid lines illustrate the rotors.}
	\label{fig:wakeVelocity}
\end{figure}

\begin{figure}
	\centering
	\includegraphics[width=0.8\textwidth]{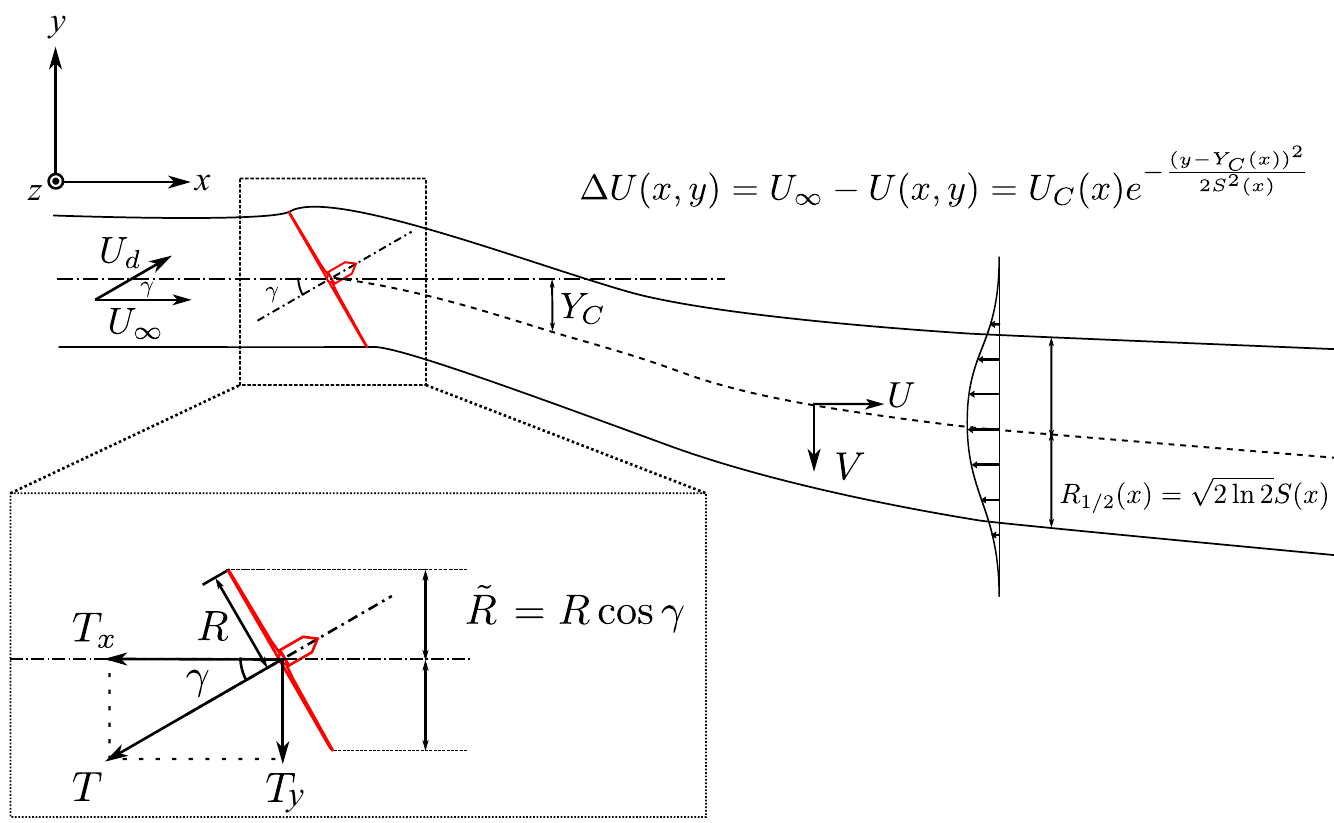}
	\caption{Schematic and and key variables for describing the wake behind a yawed wind turbine on the hub-height plane, where $U_\infty$ is the inflow velocity, $U_d$ is the inflow velocity component in the rotor's axis direction, $\gamma$ is the yaw angle, $T$ is the trust of the rotor, and $R = 0.5D$ is the rotor's radius. $U$ and $V$ denote the streamwise and the transverse velocity, respectively.}
	\label{fig:wakeSketch}
\end{figure}

In this subsection, we examine the influence of yaw angle on the time-averaged wake velocity. To illustrate such an influence, we first plot in figures \ref{fig:wakeVelocity} the streamwise and transverse velocity $U$ and $V$ 
on the hub-height plane for $\lambda = 8$ case (the rest TSR cases have the similar pattern).   The red thick dashed  line and the red dotted lines plot the wake centerline ($y = Y_C(x)$) and the boundary ($y = Y_C(x) \pm R_{1/2}(x)$) obtained by fitting the time-averaged velocity deficit using the Gaussian fit at each downstream locations ($x$), as follows, 


\begin{equation}
     \Delta U(x,y) = U_\infty - U(x,y) = \Delta U_C(x) e^{-\frac{(y-Y_C(x))^2}{2 S^2(x)}}, \label{eqn:selfsimilar}
\end{equation}
where $U(x,y)$ is streamwise velocity. A schematic of these definitions is illustrated in figure \ref{fig:wakeSketch}. The wake half-width is defined as $R_{1/2}(x) = \sqrt{2\ln 2} S(x)$, which gives the distance from $Y_C(x)$ to the position where $\Delta U = \displaystyle \frac{1}{2}\Delta U_C$. 

First, it is seen in figure 3 that the streamwise velocity $U$ 
behave similarly in yawed and non-yawed cases. 
As seen, $U$ is symmetry with respect to the wake centerline. The overall patterns of  $U$ computed at different yaw angles are very similar, although the magnitude of velocity deficit ($\Delta U$) and the wake width ($R_{1/2}$) gradually reduce with the increase of yaw angle due to the reduction of streamwise thrust of the rotor $T_x = T \cos{\gamma}$ and the projected rotor width $\widetilde{R} = R \cos{\gamma}$.  

On the contrary, the transverse velocity $V$ behaves differently in yawed and non-yawed cases. As seen in the second column of figure \ref{fig:wakeVelocity}, in the region close to the rotor, the flow at different yaw angles are all similar to the flow around a bluff-body. However, significant differences are observed in the wakes: in the non-yawed case, the transverse velocity is close to zero and is symmetrical to the wake centerline; when the rotor yaws, on the other hand, the thrust component in the transverse direction introduces transverse velocity in the negative $y$ direction with its magnitude increasing via yaw angle. More importantly, the transverse velocity is observed being asymmetric with
respect to the wake centerline $Y_C(x)$ defined with the streamwise velocity deficit. The transverse velocity  resides mainly above the wake centerline, which is in agreement with the wind tunnel measurement of \cite{bastankhah_experimental_2016}. Due to these differences observed in the transverse velocity $V$ as compared with streamwise velocity $U$, it is expected that the wake quantities defined by $U$ and $V$ should be treated differently. For this reason, we define in addition the wake center position, $Y_C^V(x)$, wake velocity on the centerline $V_C(x)$, wake width $R_{1/2}^V(x)$ for $V$ in the same way as for the streamwise velocity characteristics.  

In the following subsections, we will examine the similarity of these wake characteristics, using different sets of characteristic length and velocity scales presented in Appendix A. It will be shown that the quantities defined based on $U$, \ie{}, $\Delta U_C$ and $R_{1/2}$ are properly scaled using the set of characteristic scales derived using the streamwise thrust component ($T_x$) and the rotor area projected to the streamwise direction (based on $\widetilde{R}$). The quantities related to $V$ ( \ie{}, $Y_C(x), V_C(x), R_{1/2}^V(x)$), on the other hand, collapse with each other when scaled using characteristic scales defined using the transverse component of thrust $T_y$.

\subsection{Similarity of wake centerline $Y_C(x)$}

\begin{figure}
	\centering
	\includegraphics[width=\textwidth]{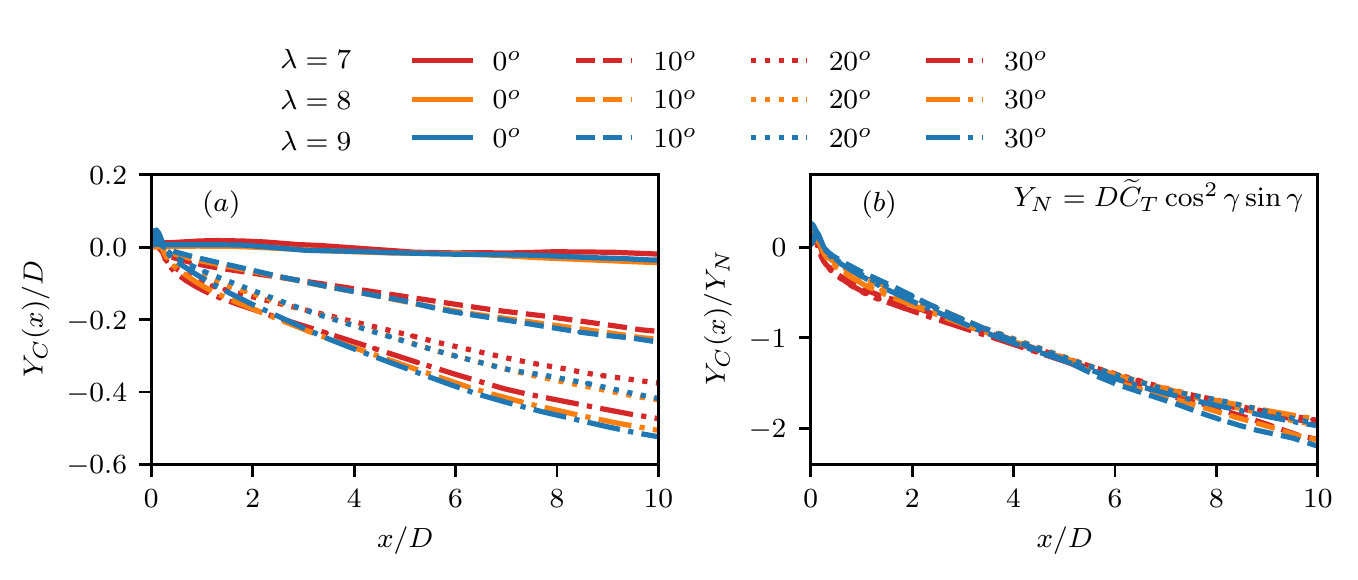}
	\caption{Wake center line $Y_C(x)$: (a) normalized by rotor's diameter $D$ (b) normalized by the length scale $Y_N = D \widetilde{C}_T\cos^2 \gamma \sin \gamma $. } \label{fig:wakeCenterLines}

	
	\label{fig:wakeCenterline}
\end{figure}

In figure \ref{fig:wakeCenterLines}, we examine the similarity of the wake centerline $Y_C(x)$ for different yaw angles. The length scale employed to normalize $Y_C(x)$ is
\begin{equation}
    Y_N =  D \widetilde{C}_T \cos^2 \gamma \sin \gamma.
    \label{eqn:wakeCenterDeflection}
\end{equation}
It is seen in figure \ref{fig:wakeCenterline} (a) that the wake centerline deflections increase with yaw angle.  
For the same yaw angle, it is observed that $Y_C(x)$ increases with TSR  at far wake locations especially for the $\gamma=30^{\circ}$ case. However the differences of $Y_C$ between cases of different TSRs are small because of the relative small difference in the thrust coefficients $C_T$ for the considered TSRs. In figure \ref{fig:wakeCenterline} (b), it is observed that the wake centerlines collapse on each other for different yaw angles after being normalized by $Y_N$. This good scaling shows that the simple length scale defined with equation \eqref{eqn:wakeCenterDeflection}  can be employed to describe the similarity of the wake centerline despite the complex dynamics of yawed turbine wakes. Once the wake centerline is known for one yaw angle, it can be generalized with the proposed length scale $Y_N$ to predict the wake centerlines at other yaw angles.

\subsection{Similarity of streamwise velocity deficit $\Delta U$}


In this section, we examine the similarity of quantities related to the streamwise velocity deficit. The characteristic velocity employed for scaling is 

\begin{equation}
    U_N = U_\infty  \left(1-\sqrt{1-\widetilde{C}_T\cos^2 \gamma }\right), \label{eqn:velocityDeficit}
\end{equation}
which represents the streamwise velocity reduction caused by the rotor. Deviation of this characteristic velocity is given in Appendix A. 

In figure \ref{fig:velocityDeficit}, we first examine the streamwise evolution of the velocity deficits  $\Delta U_C(x)$ along the wake centerline for different yaw angles and different TSRs. 
As seen in figure \ref{fig:velocityDeficit} (a), profiles from $\Delta U_C(x)$ at different TSRs and different yaw angles deviate from each other, showing the influence of both factors. For the same TSR, $\Delta U_C(x)$ decreases nonlinearly with the increase of yaw angle that the difference between $\gamma = 0$ and $\gamma = 10^o$ is smaller compared to that between $\gamma = 20^o$ and $\gamma = 30^o$. For the same yaw angle, $\Delta U_C(x)$ increases when increasing $\lambda$ due to the increase of $C_T$. More importantly, at all considered downstream locations ($1D\le x \le 10D$), $\Delta U_C(x)$ profiles are observed  varying in a similar manner.  This similarity is further confirmed in figure \ref{fig:velocityDeficit} (b), showing that all $\Delta U_C(x)$ profiles collapse with each other when normalized using the characteristic velocity $U_N$. In the figure, it is also noticed that $\Delta U_C(x)$ first increases until about $x = 2.5D$  and then gradually decreases to far wake locations. This increase of velocity deficit is related to the root loss around the hub in the near wake as shown in figure \ref{fig:wakeVelocity} and later in figure \ref{fig:velocityDeficitSelfSimilarity}, which makes the velocity deficit deviate from the Gaussian shape and may also be responsible for the slightly differences among the scaled profiles observed in the near wake. In the far wake, the scaled wake center velocity $\Delta U_C(x)/U_N$ shows an excellent agreement with each other indicating $U_N$ is the proper velocity scale for velocity deficit.   

\begin{figure}
	\centering
	\includegraphics[width=\textwidth]{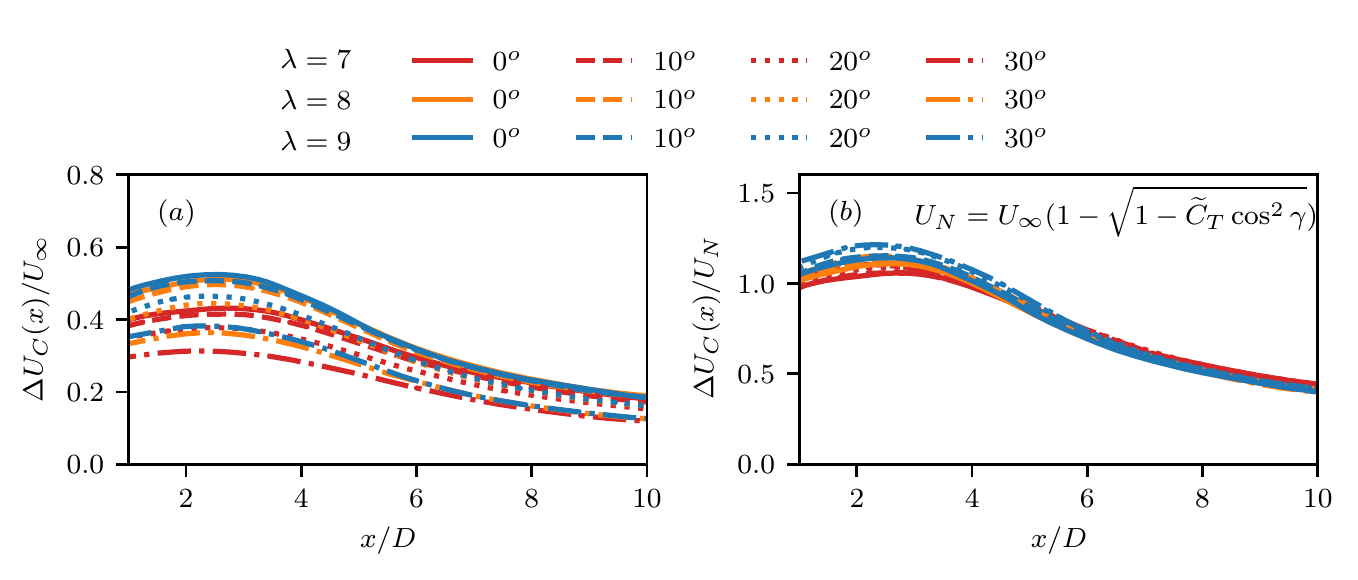}
	\caption{The characteristic velocity deficit $\Delta U_C(x)$ for different TSRs and yaw angles: (a) normalized by the inflow velocity $U$, (b) normalized by the velocity scale as $U_N = U_\infty  \left(1-\sqrt{1-\widetilde{C}_T\cos^2 \gamma }\right) $. 
	\label{fig:streamwiseVelocityVariation}} 

	\label{fig:velocityDeficit}
\end{figure}

\begin{figure}
	\centering
    \includegraphics[width=\textwidth]{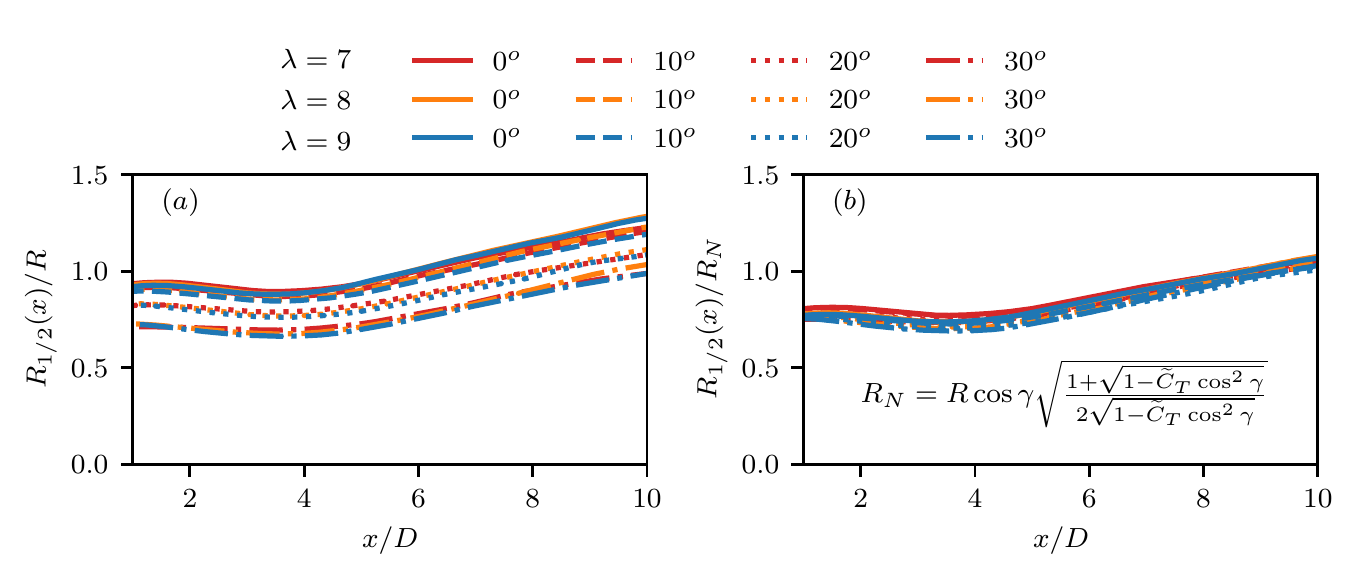}
	\caption{The characteristic wake width $R_{1/2}(x)$ for different TSRs and yaw angles: (a) normalized by the rotor radius $R$, (b) normalized by the length scale $R_N =R \cos{\gamma}\sqrt{\frac{1+\sqrt{1-\widetilde{C}_T\cos^2 \gamma}}{2\sqrt{1-\widetilde{C}_T\cos^2 \gamma}}} $.} 
	\label{fig:wakeWidth}
\end{figure}
\begin{figure}
	\centering
	\includegraphics[width=\textwidth]{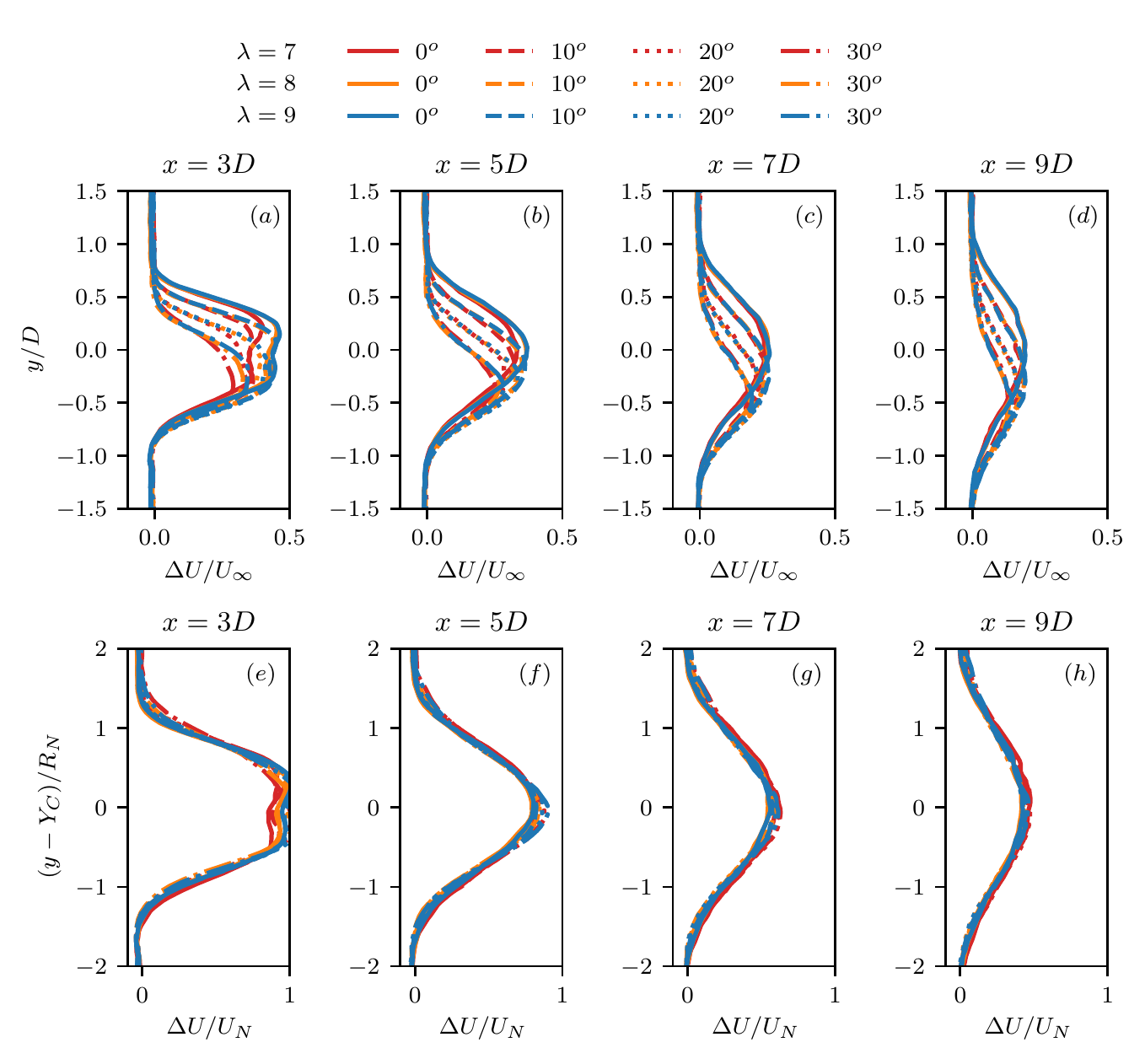}
	\caption{The time-averaged velocity deficit $\Delta U = U_\infty - U$ at downstream locations of  ($ x \in \{3D,5D,7D,9D\}$) on the hub height plane ($ z = z_\textrm{hub}$) for different TSRs and yaw angles: (a)-(d) normalized by the inflow velocity $U_\infty$, (e)-(h) the wake centers are shifted with respect to $Y_C$ with the values normalized by the proposed velocity and length scale $U_N$ and $R_N$.}  

\label{fig:velocityDeficitSelfSimilarity}
\end{figure}

We then examine the wake half-width $R_{1/2}(x) = \sqrt{2\ln{2}}S(x)$, where $S(x)$ is obtained by fitting the velocity deficit using the Gaussian distribution (equation \eqref{eqn:selfsimilar}). The initial wake width $R_N$, derived using the streamwise one-dimensional momentum  theory (details can be found in Appendix A), is employed for the normalization.  The expression of $R_N$ is shown as follows:  
\begin{equation}
    R_N = R \cos{\gamma}\sqrt{\frac{1+\sqrt{1-\widetilde{C}_T\cos^2 \gamma}}{2\sqrt{1-\widetilde{C}_T\cos^2 \gamma}}}. 
\end{equation}
which is a function of the thrust  coefficient $\widetilde{C}_T$, yaw angle  $\gamma$ and the rotor's radius $R$. In figure \ref{fig:wakeWidth}, the performance of this scaling factor is examined. We first examine the streamwise variations of wake half-width $R_{1/2}$ normalized using $R$ in figure \ref{fig:wakeWidth} (a). As seen, there are differences in $R_{1/2}/R$ among cases of different yaw angles and different TSRs. 
We then show in figure \ref{fig:wakeWidth} (b), the streamwise profiles of wake half-width $R_{1/2}$ normalized using $R_N$ for cases of different yaw angles. As seen, they collapse with each other at all considered downstream locations in the range of $1D<x<10D$, showing the proposed length scale $R_N$, a characteristic wake width scale in the near wake,  is still valid even in the far wake. It is also noticed in the figure that the wake width slightly decreases in the initial near wake region, which probably can be attributed to the root loss phenomenon shown in figure \ref{fig:wakeVelocity}. It is noted that \cite{bastankhah_experimental_2016} proposed an analytical expression for predicting the wake width at the onset of far wake assuming small $C_T$ and small $\gamma$ (see details in Appendix B). The similarity observed in figure \ref{fig:wakeWidth} demonstrates that the wake width variation at any yaw angle can be predicted using the streamwise variation of the wake width at one yaw angle and the proposed length scale in both near and far wake. 

After showing the similarity of the velocity deficit along the wake centerline and wake width. Here we examine the similarity of the transverse profiles of the streamwise velocity deficit at different turbine downstream locations in figure \ref{fig:velocityDeficitSelfSimilarity}. As shown figure \ref{fig:velocityDeficitSelfSimilarity} (a-d), the magnitude of the velocity deficit and the wake width decrease when increasing yaw angles. By contrast, transverse profiles of the streamwise velocity deficit shown figure 7 (e-h) collapse on each other when normalized using the velocity scale $U_N$ and the length scale $R_N$ and being shifted with respect to the wake centerline $Y_C(x)$. It is seen from figure \ref{fig:velocityDeficitSelfSimilarity} (e-h) that the velocity deficit profiles are symmetrical to the wake centerline for different yaw angles including $\gamma = 30^o$. The characteristic wake width scale $R_N$ and velocity deficit scale $U_N$ are proper for describing the similarity of velocity deficits for different yaw angles. 
It is noticed that at $3D$ turbine downstream, the velocity deficit  $\Delta U/ U_N $ on the plateau are approximately equal to 1, indicating that the present scaling factor $U_N$ is an accurate estimation of the velocity deficit in the near wake for different yaw angles. Furthermore, the collapse of velocity deficit profiles at further turbine downstream locations (figures \ref{fig:velocityDeficitSelfSimilarity} (f-h)) shows that $U_N$ successfully describes the similarity of velocity deficits for  different yaw angles, although it is derived at the imminent near wake. This similarity of velocity deficit profiles for different yaw angles shows the feasibility to obtain the wake velocity field behind a yawed turbine by transforming a non-yawed turbine's wake using the proposed characteristic velocity and length scales plus the wake centerline position $Y_C(x)$. 

\subsection{Transverse velocity $V$}

In this subsection, we investigate the similarity of the transverse velocity $V$. As shown in figures \ref{fig:wakeVelocity}, the transverse velocity field is significantly different from streamwise velocity and is  thus not expected to have the same characteristic velocity and length scales. Consequently, the characteristic
quantities scales defined using the transverse component of the thrust $T_y$ will be employed for scaling the quantities related to the transverse velocity, \ie{},  $R^V_{1/2}(x)$ and $V_C(x)$. 



\begin{figure}
	\centering
	\includegraphics[width=\textwidth]{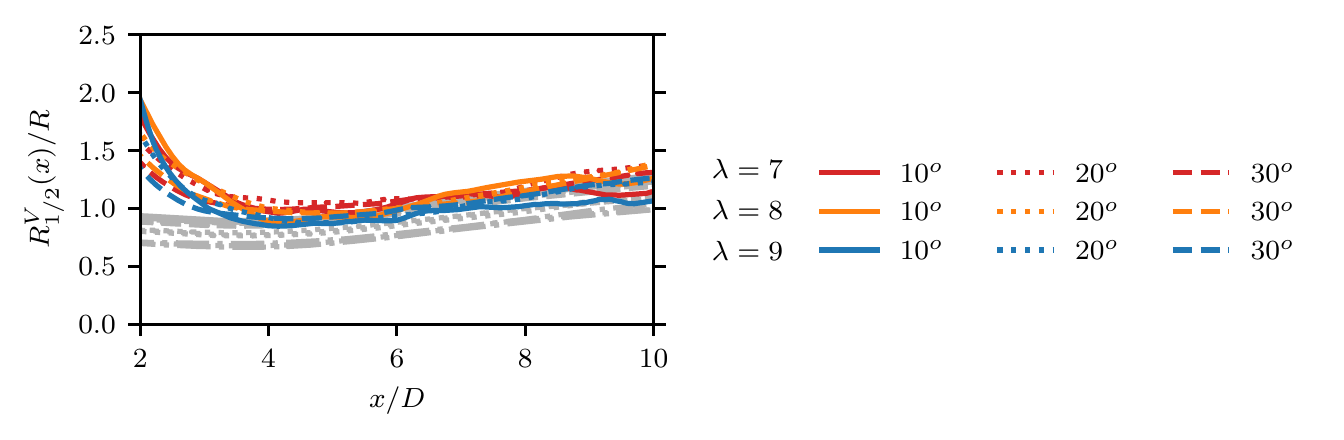}
	\caption{Streamwise variations of the half-width $R^V_{1/2}(x)$ computed by Gaussian fitting the transverse profiles of the transverse velocity $V$. The grey curves replot $R_{1/2}(x)$ computed from the streamwise velocity in figure \ref{fig:wakeWidth} (a) for reference.}  
		\label{fig:transverseVelocityWidthOverX2}

\end{figure}

We first examine the wake half-width $R_{1/2}^V(x)$ defined using the transverse profiles of the transverse velocity. 
Figure \ref{fig:transverseVelocityWidthOverX2} shows the streamwise variation of  $R^V_{1/2}$. It is seen that the streamwise $R_{1/2}^{V}$ profiles computed from cases of different yaw angles collapse on each other in the far wake  ($x>4D$) when normalized using rotor radius $R$ without introducing another characteristic length, in contrast to the half-width of the streamwise velocity deficit $R_{1/2}$, which decreases with yaw angle.  Moreover, the width $R_{1/2}^{V}$ is much wider in the near wake than in the far wake, which is different from $R_{1/2}$ and should be related to the flow around bluff-body pattern in the near wake. It is also observed that $R_{1/2}^V(x)$ decreases rapidly in the near wake for all considered cases, while $R_{1/2}(x)$ barely varies along the streamwise direction in the near wake. 
In the far wake locations ($x > 4D$), on the other hand, both  $R_{1/2}^{V}(x)$ and $R_{1/2}(x)$ gradually increase with the distance from turbine. The slope of all the curves for both $R_{1/2}(x)$ and $R_{1/2}^{V}(x)$ are close to each other in the far wake region. However, it is observed that the wake width defined using the transverse velocity $R_{1/2}^{V} $ is observed being slightly larger than $R_{1/2}$  defined using the streamwise velocity, showing a different influencing region of the transverse velocity and the wake featured by the wake deficit. This different influencing region of $V$ indicates that the transverse thrust component not only affects the flow momentum in stream tube encompassing the rotor, but also the flow around  it. This implies that it is more difficult to predict the transverse velocity with the momentum theory since its influencing zone is not clearly defined compared to that of streamwise velocity and 
special attention has to be paid when employing the width based on the streamwise velocity $R_{1/2}(x)$ to computing the far wake transverse velocity $V$ in analytical wake models for yawed wind turbines based on the momentum theory \citep{jimenez_yaw_2010,bastankhah_experimental_2016,qian_new_2018}. 

\begin{figure}
	\centering
	\includegraphics[width=\textwidth]{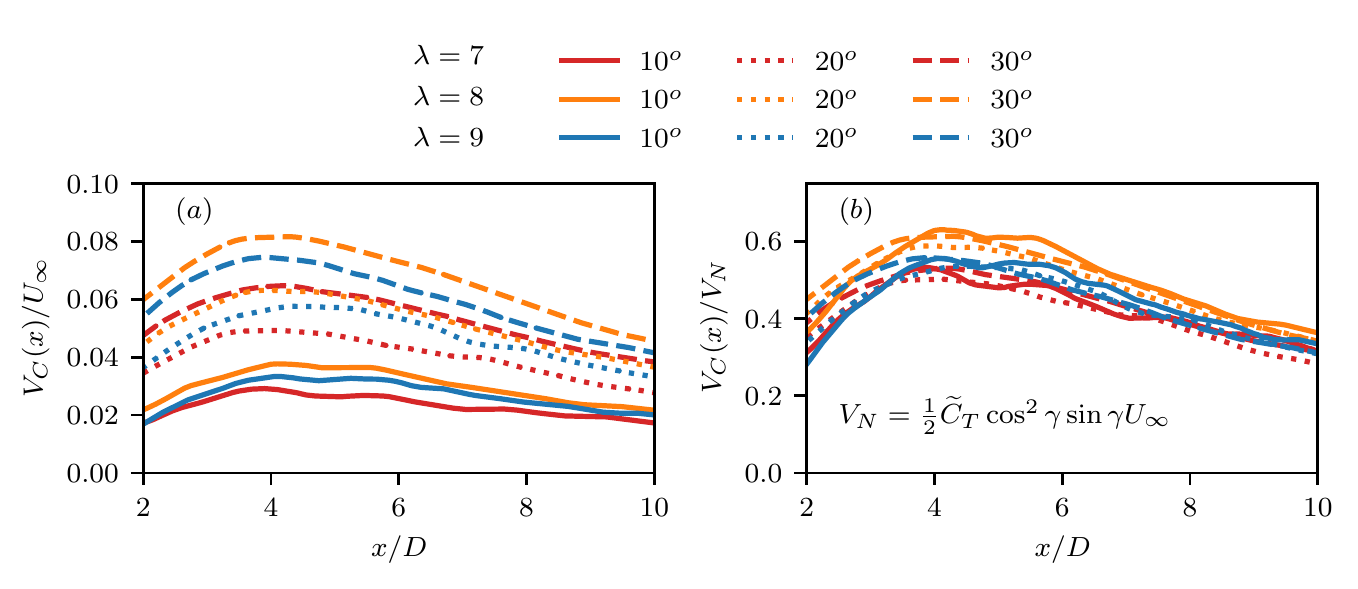}
	\caption{The characteristic velocity deficit $\Delta V_C(x)$ for different yaw angles: (a) normalized by inflow velocity $U_\infty$, (b) normalized by a the proposed velocity scale $V_N = \frac{1}{2} \widetilde{C}_T \cos^2{\gamma} \sin{\gamma} U_\infty $ . }  
	\label{fig:transverseVelocityOverX}
\end{figure}

\begin{figure}
	\centering
	\includegraphics[width=\textwidth]{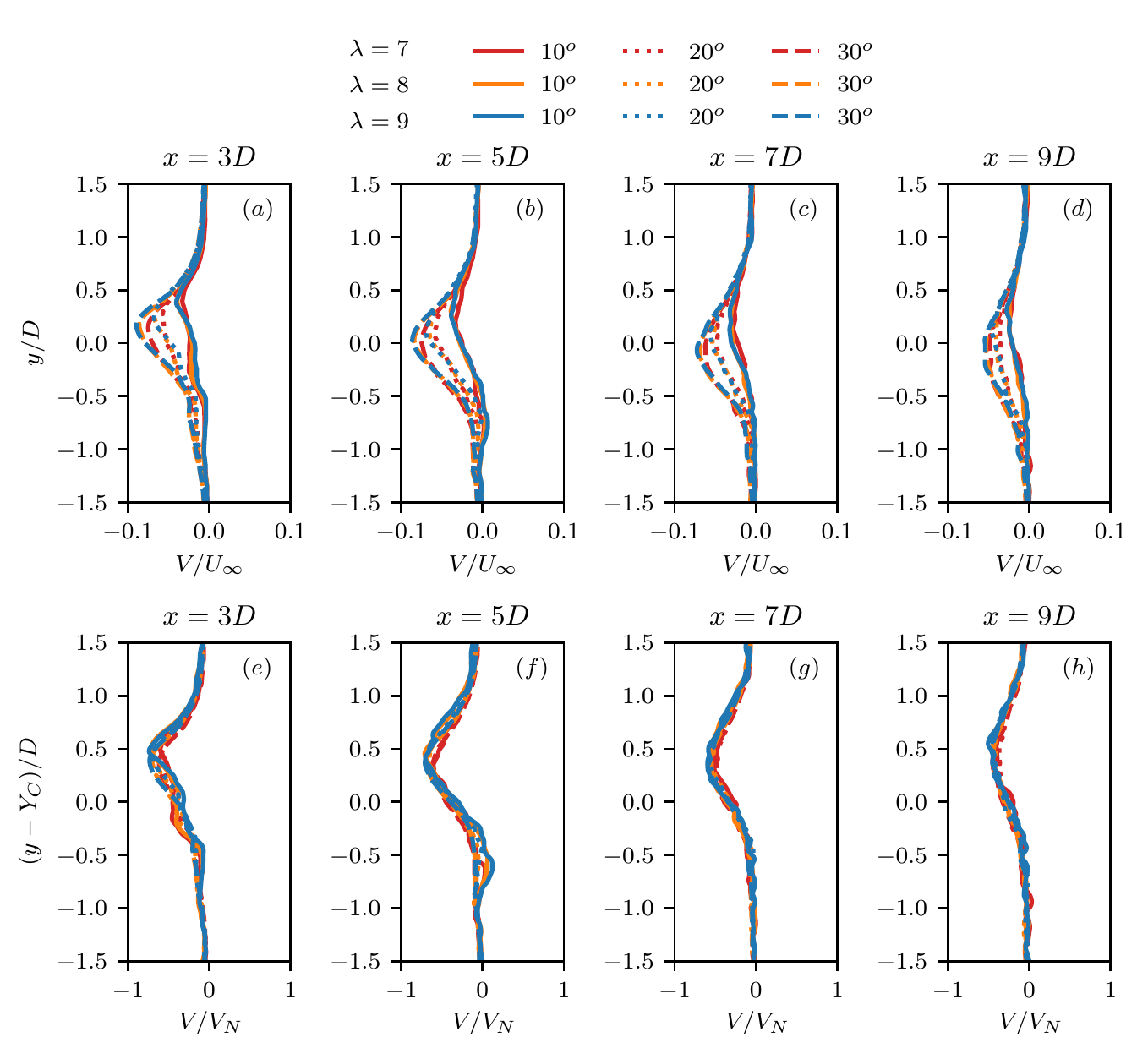}
	\caption{The time-averaged transverse velocity $V$ profile at downstream locations of  ($ x \in \{3D,5D,7D,9D\}$) on the hub height plane ($ z = z_\textrm{hub}$) for different yaw angles: (a)-(d) $V$ normalized by the inflow velocity $U_\infty$, (e)-(h) the wake centers are shifted with respect to $Y_C$ with the values normalized by the velocity scale $V_N = \frac{1}{2}\widetilde{C}_T \cos^2 \gamma \sin \gamma U_\infty$.}  
	\label{fig:transverseVelocityDeficitProfile}
\end{figure}

We then examine the similarity of the transverse velocity  using the velocity scale $V_N$ given as follows: 

\begin{equation}
V_N = \frac{1}{2}\widetilde{C}_T \cos^2 \gamma \sin \gamma U_\infty, \label{eqn:transverseVelocitySimilarity} 
\end{equation}
Figure \ref{fig:transverseVelocityOverX} shows the streamwise variations of the maximum transverse velocity $V_C(x)$ obtained from Gaussian fit of the transverse velocity. As seen in figure \ref{fig:transverseVelocityOverX} (a), significant differences are observed among cases of different yaw angles and tip-speed ratios.  Figure \ref{fig:transverseVelocityOverX} (b)  shows the downstream variations of $V_C$ normalized using the corresponding $V_N$.  As seen, $V_C(x)/V_N$ increases from approximately $0.2$ at $x = 2D$ to $0.3$ at $x=4D$ and then gradually decreases to about $0.16$ at $x = 10D$. More importantly, it is seen that the $V_C/V_N$ profiles for different yaw angles approximately collapse on each other especially at far downstream locations, showing that the similarity is captured by the characteristic velocity $V_N$. It is also found that the $V_C/V_N$ curves for different yaw angles and TSRs have slightly larger differences as compared with  $U/U_N$ in figure \ref{fig:streamwiseVelocityVariation}. This discrepancy is probably caused by the fact that the transverse profile of $V_C$ is slightly different from the Gaussian function's bell shape and there are larger errors in estimating $V_C$ by the curve fitting, as will be shown in figure \ref{fig:transverseVelocityDeficitProfile}.



With the similarity observed on $R_{1/2}^V(x)$ and $V_C(x)$, we here examine the similarity of the transverse profile at $ x \in \{3D,5D,7D,9D\}$. In figure \ref{fig:transverseVelocityDeficitProfile} (a-d), the $V$ profiles are normalized using the incoming wind speed $U_{\infty}$. As seen, for a fixed TSR, the velocity amplitude increases when increasing yaw angle for all the considered streamwise locations. 
At $3D$ turbine downstream location, the $V$ profiles show complex variations, especially for the cases with yaw angle $\gamma = 10^o$. In figure \ref{fig:transverseVelocityDeficitProfile} (e-h), the $V$ profiles are normalized with the proposed velocity scale $V_N$ and shifted with respect to the center location $Y_C$ based on the streamwise velocity deficit. As seen, profiles of the transverse velocity are approximately symmetrical about the summit, which is above the wake centerline $Y_C$. The distance between the summit and the wake centerline is approximately $0.5D$ and varies slightly in the downstream location. In the literature, this distance was found to be approximately equal to $S$ of equation \eqref{eqn:selfsimilar} in \cite{bastankhah_experimental_2016}.  It is also observed in figure \ref{fig:transverseVelocityDeficitProfile} (e-h) that the transverse velocity varies significantly in the transverse direction with the maximum magnitude of $V$  remarkably larger than that on the wake centerline ($y-Y_C=0$) at all considered streamwise locations. 
Moreover, the shape of the transverse profile of $V$ is observed deviating from the Gaussian function's bell shape,  especially at $x=3D$ and $x=5D$ locations, which affects the accuracy of the Gaussian fitting, showing potential larger errors in the Gaussian fitting in the near field.  Despite all these complexities, it is observed that the proposed velocity scale $V_N$ and $D$ can properly describe the similarity of the transverse velocity. 

\section{Similarity of turbulence statistics of yawed turbine wakes}

After investigating the similarity of time-averaged velocity fields. In this section, we examine the turbulence statistics of yawed turbine wakes, which include the turbine-added turbulence kinetic energy, turbine-added Reynolds shear stress, and the statistics of instantaneous wake positions. 

\subsection{Turbine-added turbulent kinetic energy and Reynolds shear stress}

In this subsection, we examine the similarity of the turbine-added turbulence kinetic energy $\Delta k$ and the turbine-added Reynolds shear stress $\Delta \langle u^{\prime} v^{\prime} \rangle$. The employed characteristic velocity scale is given as follows: 
\begin{equation}
    U_T = U_\infty\cos\gamma\sqrt{\widetilde{C}_T/2} \label{eqn:UT2}.
\end{equation}
\begin{figure}
	\centering
	\includegraphics[width=\textwidth]{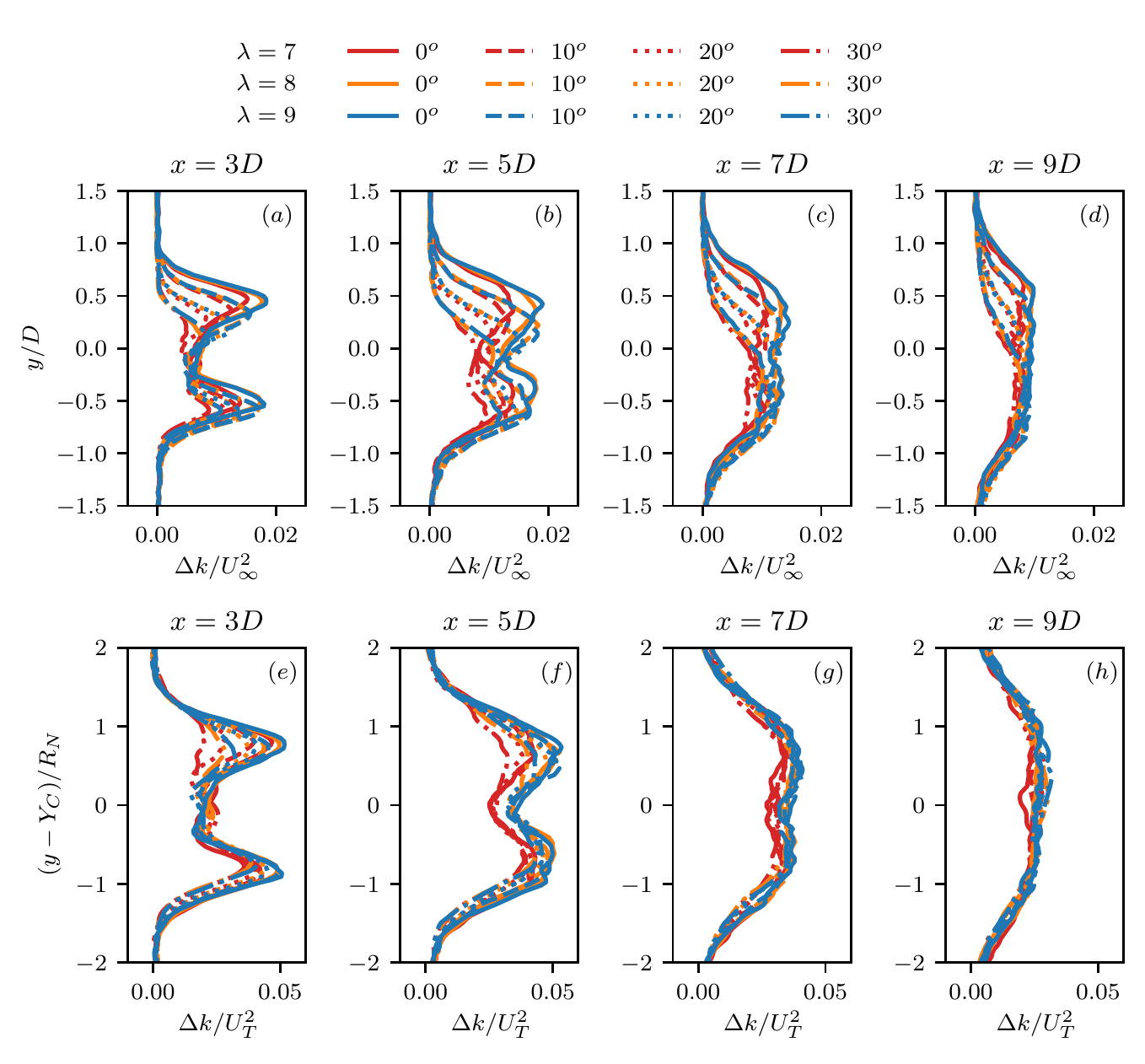}
	\caption{The turbine-added turbulence kinematic energy $\Delta k $ at downstream locations of  ($ x \in \{3D,5D,7D,9D\}$) on the hub height plane ($ z = z_\textrm{hub}$) for different yaw angles and TSRs : (a)-(d) normalized by inflow velocity square $U_\infty^2$, (e)-(h) the wake centers are shifted with respect to $Y_C$ and normalized with the length scale $R_N$, the values are normalized by the square of the velocity scale $U^2_T$.}  
	\label{fig:addedTKE}
\end{figure} 	
Figure \ref{fig:addedTKE} compares the turbine-added turbulence kinetic energy  $\Delta k$ computed from cases of different yaw angles and TSRs at different downstream locations. Plots in the first row (\ie{}, figure 12 (a-d)) are $\Delta k$ profiles normalized using $U^2_{\infty}$. As seen, $\Delta k$ profiles curves of different yaw angles show apparent differences in terms of the magnitude, the locations for $\Delta k$ peaks, and the width of the zone with increased turbulence. Figures in the second row 
show the same results with the abscissa scaled by $U^2_T$ and the ordinate shifted with $Y_C(x)$ then scaled by $R_N$. With this scaling,  the $\Delta k$ profiles of different yaw angles collapse on each other in the far wake, showing the characteristic velocity $U_T$ is still the proper velocity scale for the wake of yawed wind turbines. Further probing into the $\Delta k$ profiles in the near wake, it is found that the profile of $\Delta k$ contains two peaks with the distance between the two peaks scaled well by $R_N$ as the location of $\Delta k$ peaks appear at  $Y_C \pm S $, being consistent with results in the literature \citep{schottler_wind_2018}. It is also observed that $\Delta k$ in the near wake ($x = 3D$, figure \ref{fig:addedTKE} (a)(e)) is asymmetric behind yawed wind turbines ( especially for $\gamma = 30^o $) with the $\Delta k$ behind the trailing half of the rotor ($y-y_C<0$) larger than the leading half ($y-y_C>0$), which is similar to the flow passes an inclined circular disc \citep{calvert_inclinedDisk_1967,gao_inclinedDisk_2018}. This asymmetry increases with yaw angle and cannot be captured by the proposed velocity scale $U_T$. However, it is observed that this asymmetry only manifests in the near wake and may be considered as immaterial in utility-scale wind farm, where the turbine spacing is often larger than $5D$ \citep{mechali2006wake}. At far wake locations, this asymmetry disappear with the $\Delta k/U_T^2$ profiles collapsed well each other on the $(y-Y_C)/R_N$ coordinate.

\begin{figure}
	\includegraphics[width=\textwidth]{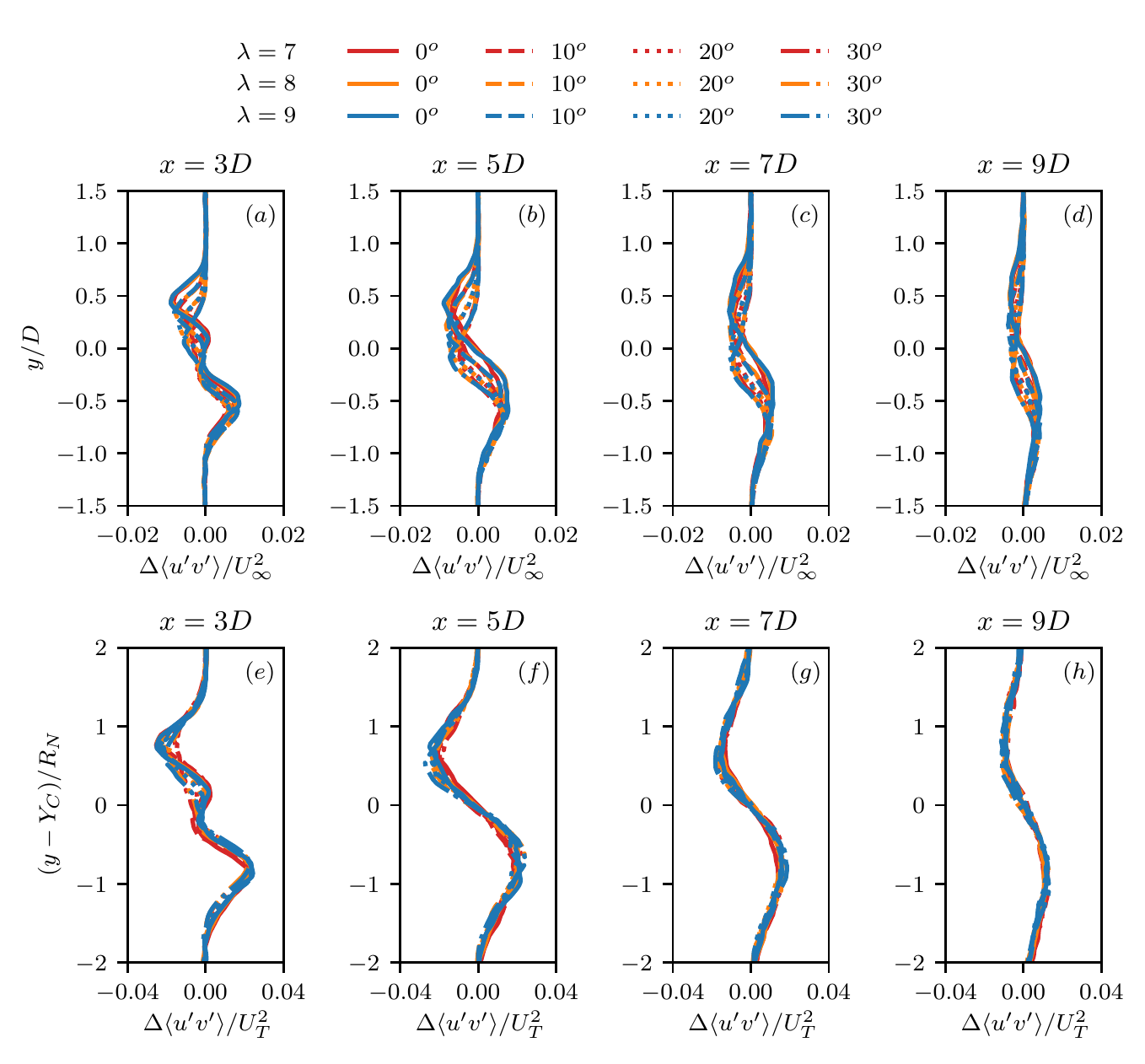}
	\caption{The turbine-added Reynolds shear stress $\Delta \langle u'v' \rangle $ at downstream locations of  ($ x \in \{3D,5D,7D,9D\}$) on the hub height plane ($ z = z_\textrm{hub}$) for different yaw angles: (a)-(d) normalised by inflow velocity square $U_\infty^2$, (e)-(h) the wake centers are shifted with respect to $Y_C$ and normalized with the length scale $R_N$, the values are normalized by the square of the velocity scale $U^2_T$.}  
	\label{fig:addedUV}
\end{figure}

Figure \ref{fig:addedUV} plots the turbine-added Reynolds shear stress $\Delta \langle u^{\prime} v^{\prime} \rangle$ at hub height plane ($z = z_\textrm{hub}$) presented in the same manner as figure \ref{fig:addedTKE}. In figure \ref{fig:addedUV} (a), it is observed that  $\Delta \langle u^{\prime} v^{\prime} \rangle$ profiles are featured by complex variations near the wake center at $x=3D$, which is caused by the turbine nacelle and root loss near the hub. As the wake develops downstream, the  $\Delta \langle u^{\prime} v^{\prime} \rangle$ profiles become asymmetrical with respect to the wake centerline, which deflects towards the negative $y$ direction when increasing yaw angle. When $\Delta \langle u^{\prime} v^{\prime} \rangle$ are plotted on the coordinate $(y-Y_C)/R_N$, the turbine-added Reynolds shear stress profiles normalized using $U_T^2$ collapse on each other in the far wake $(x \ge 5D)$  for different yaw angles and TSRs, confirming $U_T$ is the proper velocity scale for the turbine-added  Reynolds stress.

\begin{figure}
	\includegraphics[width=\textwidth]{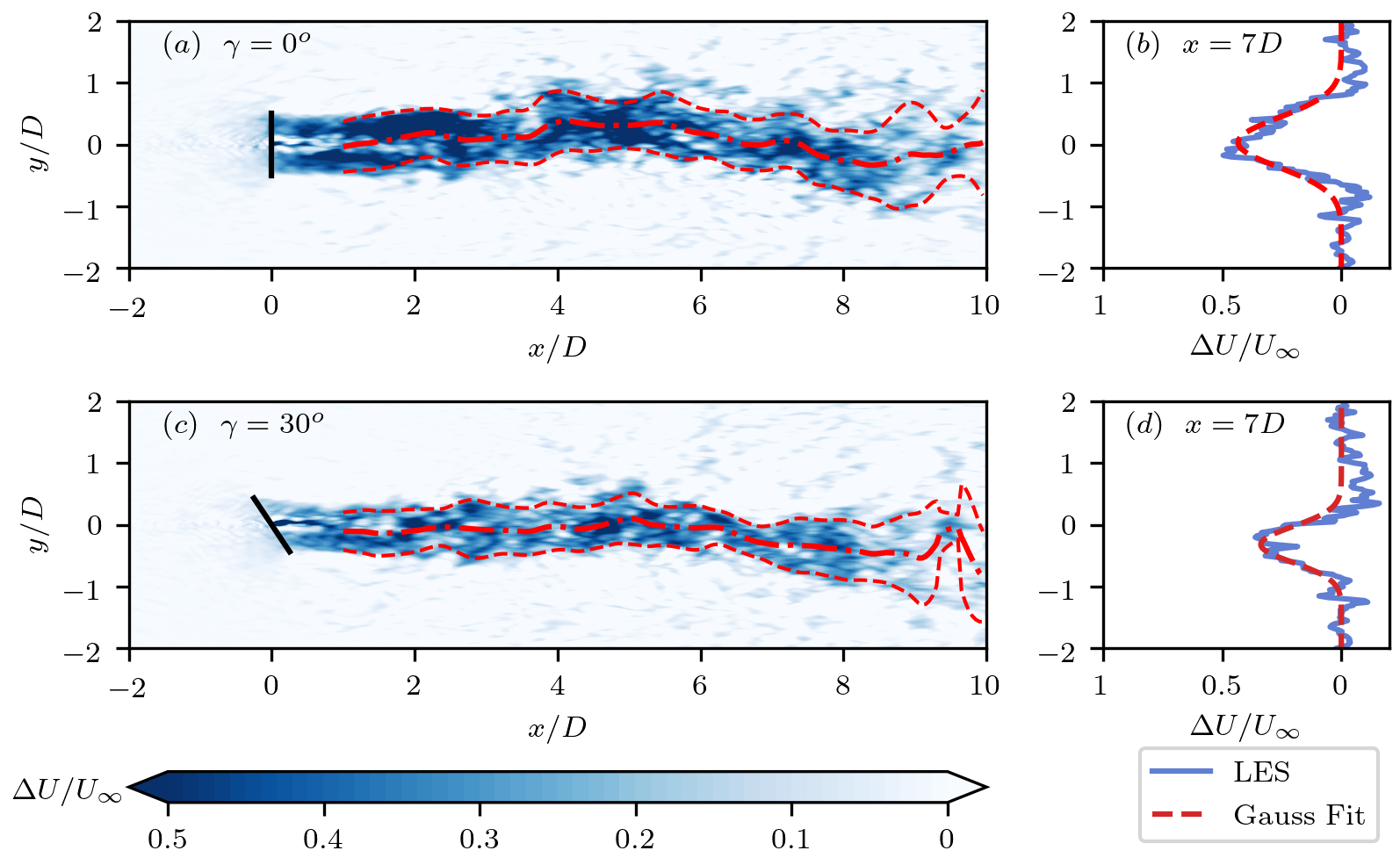}
	\caption{Instantaneous flow field behind the wind turbines at hub height plane $z = z_\textrm{hub}$. The first column shows the contour of the velocity deficit $\Delta U$ at the same simulation time for yaw angles $\gamma = 0^o$ in panel (a) and $\gamma = 30^o$ in panel (c). The black solid lines illustrates the wind turbines. Red dash-dotted lines denote the wake center line $y_c(x)$ and the red dashed lines denote the wake width defined with $r_{1/2}(x)$ obtained from Gauss fit of the instantaneous streamwise velocity. The second column plots the velocity deficit at $x = 7D$ and the corresponding Gauss fit for yaw angles $\gamma = 0^o$ in panel (b) and $\gamma = 30^o$ in panel (d).}  
	\label{fig:instantWake}
\end{figure}

\subsection{Statistics of instantaneous wake center and width}

After investigating the similarity of wake turbulence, here we analyze the influence of the yaw misalignment on the dynamics of the large-scale wake motion on the horizontal plane, which is referred to as wake meandering having significant impact on the wake expansion, recovery, and the fatigue loads on downstream wind turbines \citep{ainslie1988Meandering,hogstrom1988Meandering,larsen2007Meandering}. Specifically, we investigate the statistics of  the instantaneous wake centerline positions, which is often employed to indicate the meandering  motion of the wakes. In the same way as for the time-averaged one, the instantaneous wake center is defined as the center of the Gaussian fit of the instantaneous velocity deficit, which is obtained by subtracting the velocity field computed from the case without wind turbines at exactly the same instant. 
Figure \ref{fig:instantWake} shows the velocity deficit behind a non-yawed and yawed wind turbines at the same instant.   To obtain wake center position at each downstream location, the velocity deficit in the streamwise $\Delta U$ is first spatially filtered with filter width $0.5D$ in the streamwise direction ($x$)  and this filtered velocity profile is fitted by a Gaussian curve in the transverse direction at each downstream locations with equation \eqref{eqn:selfsimilar}.  As illustrated in figure \ref{fig:instantWake} (b) for $\gamma = 0$ and (d) for $\gamma  = 30^o$ both for $\lambda = 7 $, there are some fluctuations in the instantaneous velocity profiles (blue lines) compared with the time-averaged ones (see figure \ref{fig:velocityDeficitSelfSimilarity}). However, the fitted Gaussian curves (red dashed lines) still capture well the essential wake characteristics. Therefore the instantaneous wake center $y_c$ and the wake half-width $r_{1/2}$ can be defined from this Gaussian fit. In figures \ref{fig:instantWake} (a) and (c), the fitted wake center lines $y_c(x)$ and the characteristic wake half-width (defined with $r_{1/2}(x)$) are plotted with red dotted lines and red dashed lines, respectively. Generally, they provide a good estimation of the overall trend of the wake in the entire region except for $9D<x<10D$, where some difficulties in fitting profiles of small velocity deficit using the Gaussian function are observed. From figures \ref{fig:instantWake} (a) and (c), it is found that the wake behind a yawed wind turbine is generally narrower with smaller velocity deficit than the wake behind a non-yawed turbine. This observation is in accordance with the time-averaged wake quantities, and we will examine whether the proposed scales are still proper for the statistics of instantaneous wakes. 

The analyses in this section are based on the data collated on the horizontal plane located at turbine hub height as shown in figure \ref{fig:instantWake}. In total, 40,000 snapshots of the instantaneous velocity field are saved during the entire simulation time ($\approx$ 75 mins) at approximately 40 times of the rotor frequency.


\begin{figure}
	\includegraphics[width=\textwidth]{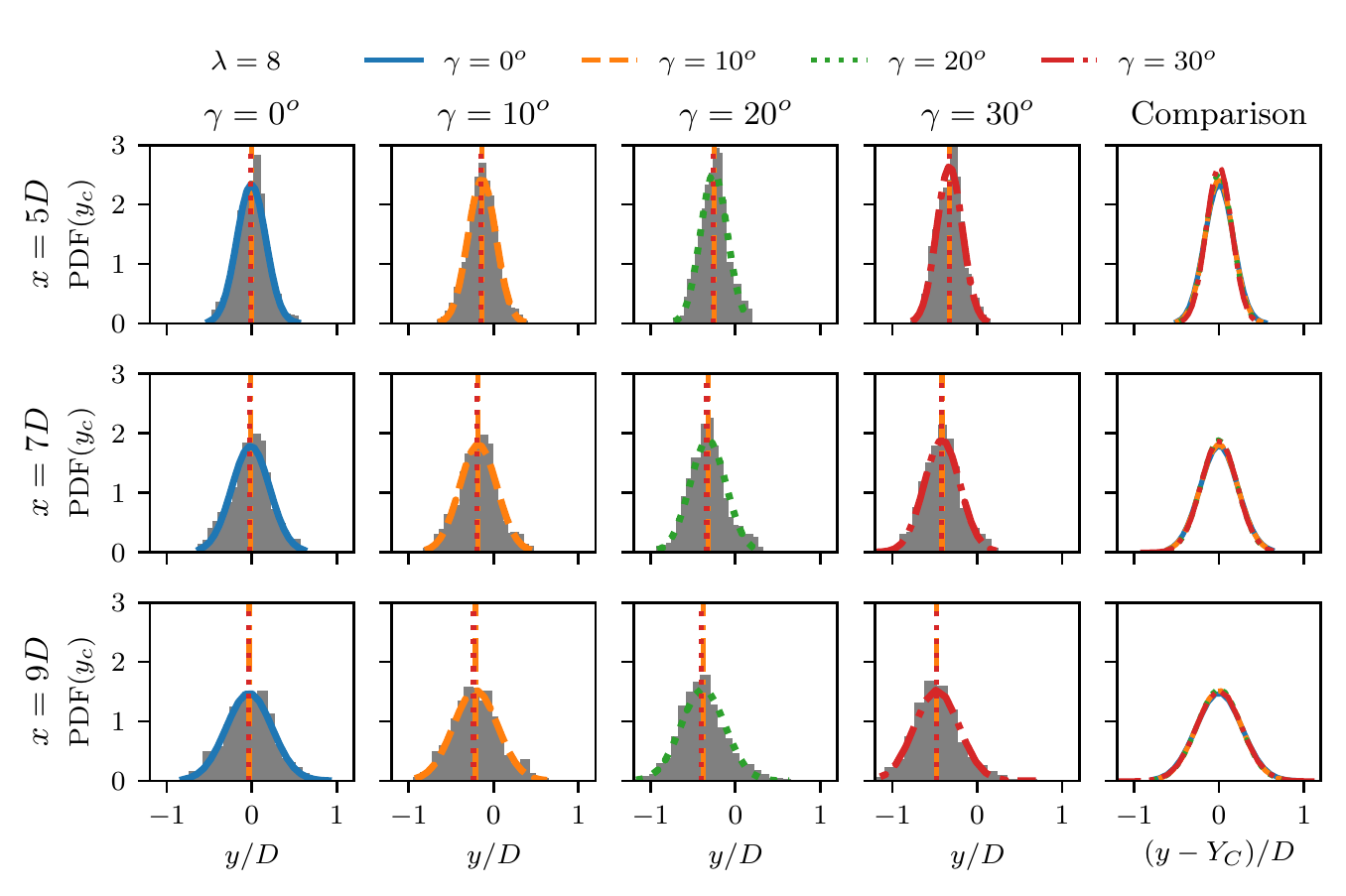}
	\caption{The probability density function of the instantaneous wake center at $x \in \{5D,7D,9D\}$ for different yaw angles at $\lambda = 8$. The histogram is fitted with Gaussian distribution function in each panel, and the fitted curves at a same down stream distance is compared in the last column. The orange dashed vertical lines denote the mean value of the instantaneous wake center locations and the red dotted vertical lines plot at wake center obtained from the time-averaged velocity as shown in figure \ref{fig:wakeCenterLines}.}  
	\label{fig:wakeCenterPDF}
\end{figure}

We first show the probability density  function (PDF) of the wake center location at $x \in \{5D,7D,9D\}$ for different yaw angles in figure \ref{fig:wakeCenterPDF}. In this figure, panels in the same rows are at the same downstream location and the panels in the same columns are at the same yaw angle. For brevity, only the results of $\lambda = 8$ are plotted as the results of all considered TSRs are similar. The PDF is plotted as grey bars and is fitted with normal distribution curves. In each panel, the two vertical lines denote the wake centerline location obtained from the time-averaged velocity field ($Y_C$) and the mean value of the instantaneous center location ($\overline{y_c}$), respectively. It is found that the wake center locations obtained from the two approaches collapse.  The PDF is found being symmetrical about the centerline and can be well approximated by the normal distribution. When increasing yaw angle, the PDF moves in the $-y$ direction due to the wake deflection. In the last column, the PDF profiles for different yaw angles are shifted with respect to $Y_C$ and plotted in the same figure for comparison. As seen, the PDF profiles of instantaneous wake positions from cases of different yaw angles collapse. This indicates that the transverse distribution of the instantaneous wake centers is independent of the yaw angle at different downstream locations for the present cases, which indicates that the amplitude of wake meandering, defined as the standard deviation of the wake center location, does not change with yaw angle. By comparing the PDFs at different locations, it is found that the wake centers distribute in a wider range as the wake travels downstream, which is in accordance with the wake expansion phenomenon observed in the instantaneous flow field. Quantitative analysis in the following sections will reveal more information about the expansion rate. 




After investigating the PDF of instantaneous wake center positions at 3 specific downstream locations in figure \ref{fig:wakeCenterPDF}, we further analyze the streamwise evolution of instantaneous wake characteristics at different yaw angles and TSRs, including both instantaneous wake center position ${y_c(x)}$ and instantaneous wake width $r_{1/2}(x)$.

\begin{figure}
	\includegraphics[width=\textwidth]{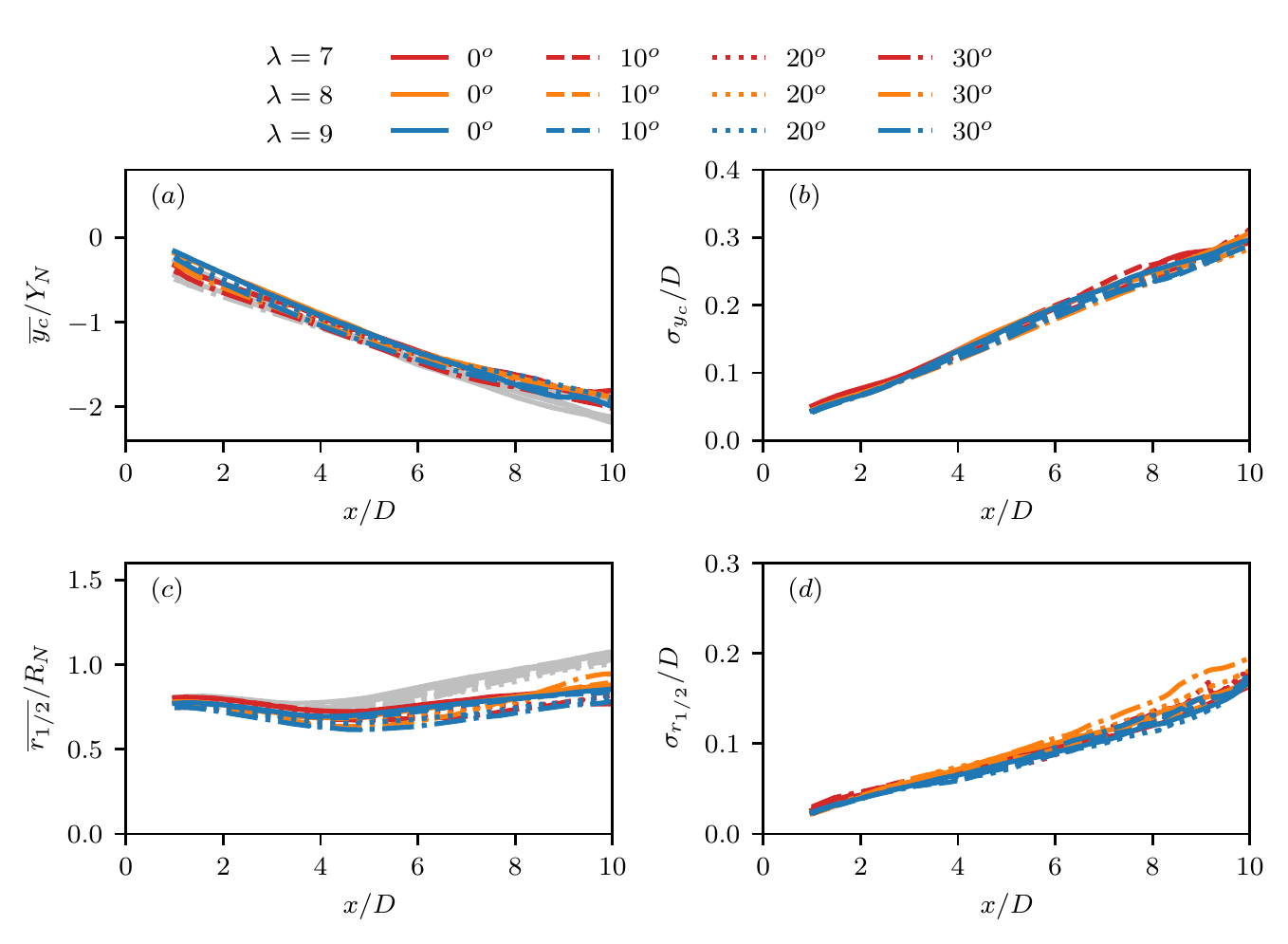}
	\caption{Streamwise distribution of the statistics of instantaneous quantities of the wakes behind yawed wind turbines, values are scaled with the same factors as for the time-averaged quantities: (a) the mean value of instantaneous wake center positions $y_c(x)$, scaled by $Y_N$; (b) The standard deviation of $y_c(x)$ normalized the rotor diameter $D$; (c) the mean value of of the characteristic instantaneous wake width $r_{1/2}(x)$ scaled by $R_N$; (d) the standard deviation of $r_{1/2}(x)$ normalized by $D$. In (a) and (c), the grey curves plots the corresponding normalized characteristics obtained from the time-averaged velocity field.}  
	\label{fig:instantQuantities}

\end{figure}

In Figure \ref{fig:instantQuantities}, we show both the time-averaged value and the standard deviation of the quantities of the instantaneous wake center and the wake width. In figures \ref{fig:instantQuantities} (a,c),  the counterparts of these values obtained from the time-averaged velocity fields are also plotted in grey scale for reference. In figure \ref{fig:instantQuantities} (a), the streamwise variations of $\overline{y_c}$ computed from different yaw angles and TSRs are compared. As seen, the mean of instantaneous of the wake center locations are well scaled using the proposed length scale $Y_N$ and collapse with the wake center obtained from the time-averaged field $Y_C$. Figure \ref{fig:instantQuantities} (b) plots the streamwise evolution of the 
standard variation of the wake center position $\sigma_{y_c}$, which indicates the amplitude of the 
meandering motion of the wake center around its mean position. In figure \ref{fig:instantQuantities} (b),  the rotor diameter $D$, instead of $Y_N$, is used for the normalization. As seen,  $\sigma_{y_c}$ from cases of different yaw angles collapse on each other. This observation shows that the wake meandering amplitudes are independent of yaw angle and TSR, which implies that the meandering motion of the wake is mainly driven by incoming large eddies for the present cases, instead of turbine operation.  
Furthermore, it is observed that $\sigma_{y_c}$ increases approximately linearly with the distance to the turbine, showing the meandering amplitude increases as wakes travel downstream. 

After investigating the wake center $y_c$, the instantaneous wake width $r_{1/2}$ is analyzed in figure \ref{fig:instantQuantities} (c) and (d). From figure \ref{fig:instantQuantities} (c), it is found that the mean value of the instantaneous wake width $\overline{r_{1/2}(x)}$ for different yaw angles and TSRs can be properly scaled by the proposed length scale $R_N$. When comparing with time-averaged wake width $R_{1/2}(x)$ shown in grey, it is found that the width defined by both approaches agree well in the near wake ($x=1D$), then $R_{1/2}$ becomes larger at further downstream locations. This is because that $R_{1/2}$, defined using the time averaged velocity field, is affected by not only the expansion of the wake around the instantaneous wake center, but also the wake meandering, which smears the velocity deficit in a wider region; in contrast, $r_{1/2}$ denotes only the width of the instantaneous wake following the instantaneous wake center $y_c$. It is also observed that the difference between $r_{1/2}$ and $R_{1/2}$ increases with turbine downstream distance in accordance with the meandering amplitude represented by $\sigma_{y_c}$.  Figure \ref{fig:instantQuantities} (d) plots the standard deviation of the instantaneous wake half-width $\sigma_{r_{1/2}}$, which represents the wake deformation that expands and shrinks with time. It is noticed that the standard deviation of the wake width for different yaw angles and TSRs collapse on each other when normalized by the rotor diameter $D$, which suggests again that the wake deformation is mainly driven by the inflow in the present cases. Furthermore, the standard deviation of $r_{1/2}$ gradually increase as traveling in the streamwise direction, indicating larger wake deformation at far wake locations. However, such phenomena have not been properly taken into account in existing wake meandering models (\eg{}, \citep{larsen2007Meandering}) and needs to be properly modeled in future development. 


In summary, figure \ref{fig:instantQuantities} mainly reveals that the mean of the instantaneous wake quantities and their standard deviation are scaled differently. The proposed scaling factors derived for the time averaged velocity field still work for the mean value of $y_c$, $r_{1/2}$, but the standard deviation of these quantities, on the other hand are independent of the length scale defined using wind turbine operation.  This independence implies the both meandering amplitude and the wake deformation is mainly driven by the inflow turbulence  for the considered cases.

\section{Summary and Conclusion}\label{sec:conclusion}

We investigate the wake characteristics of a yawed utility-scaled wind turbine under fully developed turbulent inflow using LES with blades and nacelle parameterized using the actuator surface model. Four yaw angles ($\gamma = 0^o, 10^o, 20^o, 30^o $) are considered with turbine operating at three different tip-speed ratios ($\lambda = 7,8,9$). All the cases are simulated with the same turbulent inflow obtained from a precursory LES.

The most important finding from this work is the similarity observed in  the turbine's wakes of different yaw angles for different TSRs for the time-averaged flow velocity, the turbine-added turbulence, and the statistics of the instantaneous wake center position and radius. For the time-averaged flow field, we examine two kinds of similarities, \ie{}, the similarity of the streamwise velocity deficit and the similarity of the transverse velocity. It is observed that the wake deficits $\Delta U$ from cases of different yaw angles collapse well on each other when normalized using the characteristic velocity $U_N$,  \ie{}, the velocity difference between the incoming velocity and velocity in the near wake of the turbine, and the characteristic length $R_N$,  \ie{}, the radius of the near wake, which are derived from the one-dimensional momentum theory. For the transverse motion of the wake, the transverse deflection $Y_C$ and the magnitude of the transverse velocity $V$ are observed being scaled well by the length scale $Y_N$ and the velocity scale $V_N$, respectively, which are derived using the transverse component of the thrust on turbine. The width $R_{1/2}^{V}$ of the region dominated by the transverse motion of wake, on the other hand, is similar for cases of different yaw angles and scaled well by the rotor diameter. For the turbulence characteristics of the wake, the computed results show a good scaling of the turbine-added turbulence kinetic energy $\Delta k$ and the Reynolds shear stress $\Delta \langle u' v' \rangle$ at far wake locations when normalized using $U_T$, which is defined using the streamwise component of the thrust on turbine. For the statistics of the instantaneous wake, the PDF profiles of the instantaneous wake positions $y_c$ normalized using rotor diameter are observed being collapsed on each other for cases of different yaw angles. Meanwhile, we find that the standard deviation of the instantaneous wake deflection fluctuations $\sigma_{y_c}$, of the instantaneous wake radius fluctuations $\sigma_{r_{1/2}}$ collapse on each other when normalized using the incoming velocity and rotor diameter without using the characteristic scales depending yaw angles. Overall, we have observed that the characteristics of time-averaged wake including $\Delta U$,  $R_{1/2}$, $Y_C$ and $V$ except for $R_{1/2}^{V}$ are scaled well using the velocity and length scale defined based on turbine operational conditions, while the characteristics of wake fluctuations including $PDF(y_c)$, $\sigma_{y_c}$, $\sigma_{r_{1/2}}$ are independent of yaw angles and tip-speed ratios. This suggests that yaw of turbine mainly influence the time-averaged wake characteristics, while the characteristics of wake fluctuations are largely affected by the incoming turbulence for the considered cases.  

The similarity observed on the time-averaged flow fields suggests that the wake behind a yawed wind turbine can be decomposed into a straight wake behind an equivalent non-yawed wind-turbine and a deflected centerline caused by the transverse component of the thrust on turbine, which is also the assumption for deriving the corresponding velocity and length scales. Furthermore, the simulation results show that the transverse motion and the streamwise velocity deficit of the wake resides in different regions with different influencing zones that (i) the region with strong transverse motion is found being wider than that of streamwise velocity deficit and located closer to the rotor centerline, and (ii) the wake width defined based on the streamwise velocity deficit $R_{1/2}$ decreases when increasing the yaw angle $\gamma$, while the width defined based on the transverse velocity $R^V_{1/2}$ barely changes via the yaw angle. The decomposition of the wake into streamwise and transverse components can potentially be employed to simplify the development of analytical models. The differences between the transverse velocity and the streamwise velocity deficit, on the other hand, make it difficult to predict quantities related to wake's transverse motion with analytical models based on the momentum theory and needs to be considered to improve existing analytical models \citep{jimenez_yaw_2010,bastankhah_experimental_2016,qian_new_2018}. The similarity observed in this work provides a new way to model the transverse motion of the wake from a yawed wind turbine that the wake deflection, transverse velocity and other quantities can be computed using the simulation results at one yaw angle and the velocity and length scales employed in this work.    

The observed different similarity characteristics for different turbulence statistics shows a potential for the decoupling between the meandering motion of the wake and the turbine-added wake turbulence. That the $PDF(y_c)$,  $\sigma_{y_c}$ and $\sigma_{r_{1/2}}$ scaled by $D$ suggests that wake meandering is mainly dominated by the incoming large eddies instead of the turbine operational conditions. 
The turbine-added turbulence kinetic energy can be decomposed into two parts, \ie{}, that caused by the shear layer in the wake, and the nominal part caused by the meandering motion of the wake. The meandering motion is similar for the simulated cases, such that the collapse of the $\Delta k$ profile when scaled by $U_T$ implies its similarity on the coordinate system following the meandering motion of the wake. This suggests an efficient approach to develop advanced models for wake dynamics, that the wake turbulence and the meandering motion can be modeled separately. However, further work has to be carried out to develop such models. One problem is how to model the effects of meandering motion on wake turbulence in the coordinate system following the meandering motion of the wake. To solve this problem, one approach is to treat the velocity deficit and wake turbulence as passive scales as done in the dynamic wake meandering model developed at Technical University of Denmark \citep{larsen2007Meandering}. The other problem is how to account for the bluff body shear layer instability on the meandering motion of turbine wakes \citep{heisel2018spectral,yang2019meanderingReview}, which often happens at higher frequency \citep{yang2019wake} and has been shown playing an important role in the wake of utility-scale wind turbines \citep{foti2018similarity}  and wind farms \citep{foti2019effect}.  Moreover, how incoming turbulence and stratification conditions affect the similarity observed in this work need to be investigated in the future work. 

\section*{Acknowledgement}

This work is partially supported by NSFC Basic Science Center Program for ``Multiscale Problems in Nonlinear Mechanics" (NO. 11988102 ).

\section*{Declaration of Interests}

The authors report no conflict of interest.

\appendix

\section{Derivation of velocity scales and length scales for yawed turbine wakes}

\begin{figure}
	\includegraphics[width=\textwidth]{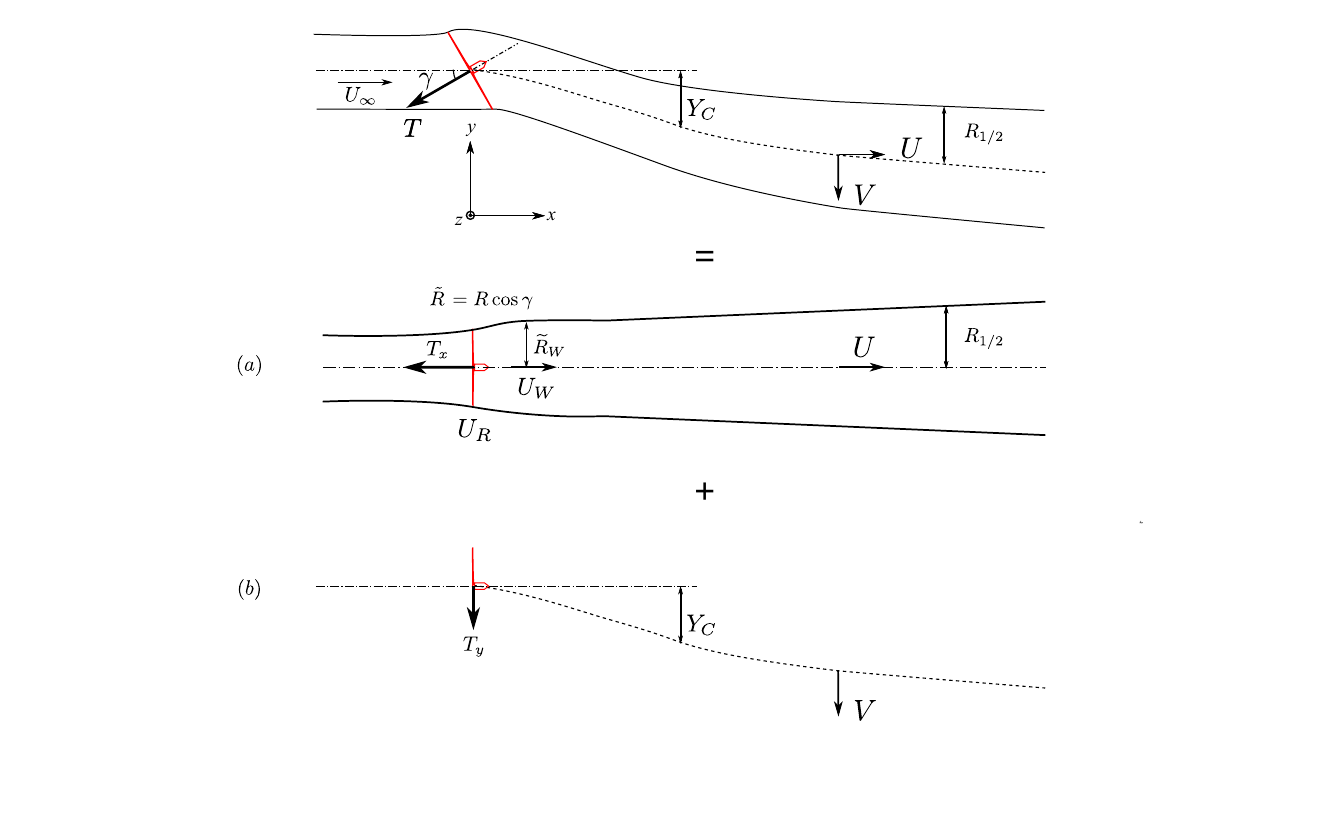}
	\caption{Decomposition of the wake behind a yawed wind turbine into: (a) an equivalent non-yawed wind turbine's wake of streamwise thrust $T_x$ and (b) the wake transverse velocity and the centerline deflection caused by the transverse thrust component $T_y$.}  
	\label{fig:wakeDecomposition}
\end{figure}

In this appendix we derive the velocity and length scales for yawed turbine wakes based on one-dimensional momentum theory. We assume that the streamwise velocity $U$ and the spanwise velocity $V$ are independent and the wake can be decomposed into a straight wake generated by an equivalent non-yawed turbine and wake deflection due to yaw of the turbine as illustrated in figure \ref{fig:wakeDecomposition}.

First, we derive the characteristic velocity $U_N$ and characteristic length $R_N$ by applying one-dimensional momentum theory to the equivalent non-yawed turbine. The equivalent non-yawed turbine is defined as a turbine of thrust $T_x$ the same as the streamwise component of the yawed turbine and and an elliptical rotor-swept area with the horizontal radius being $\tilde{R} = R \cos \gamma$ and the vertical radius being $R$ and computed as
\begin{equation}
    \tilde{A} = \pi \tilde{R} R =  \pi R^2 \cos \gamma =A \cos \gamma
\end{equation}
with $A$ the rotor sweeping area. With the thrust component aligned with the inflow computed as $T_x = T \cos \gamma$,  the thrust coefficient of this equivalent turbine can be calculated as, 
\begin{equation}
    C_{T_x} = \frac{T_x}{\frac{1}{2} \tilde{A} \rho U_\infty^2} = \frac{T \cos \gamma }{\frac{1}{2} A  \cos \gamma \rho U_\infty^2} =  \frac{T  }{\frac{1}{2} A \rho  U_\infty^2}. \label{eqn:CtOfProjectedRotor}
\end{equation}
Substituting equation \eqref{eqn:thrust} into equation \eqref{eqn:CtOfProjectedRotor} yields, 
\begin{equation}
    C_{T_x}  = \widetilde{C}_T \cos^2 \gamma.   
\end{equation}
Based on the one-dimensional momentum theory \citep{burton2011wind}, the streamwise velocity $U_W$  in the turbine's near wake is obtained as follows:  
\begin{equation}
    U_W = U_\infty\sqrt{1- C_{T_x}}  = U_\infty \sqrt{1-\widetilde{C}_T\cos^2\gamma}. \label{eqn:uwake}   
\end{equation} 
$U_N$ is then obtained as the different between the inflow velocity $U_\infty$ and the wake velocity $U$ as in equation \eqref{eqn:velocityDeficit}, as follows, 
\begin{equation}
    U_N = U_\infty - U_W =  U_\infty\left(1-\sqrt{1-\widetilde{C}_T\cos^2\gamma}\right),  
\end{equation}
which is the streamwise velocity deficit in the near wake of the equivalent non-yawed wind turbine. 

The length scale for the wake width $R_N$ is defined as the width of the imminent wake of the equivalent non-yawed wind turbine, which reflects the expansion of the stream-tube encompassing the rotor due to the flow deceleration and can be computed from the mass conservation equation.  To derive $R_N$, we first compute the streamwise velocity at the rotor disc, as 
\begin{equation}
    U_R = \frac{1}{2}\left(U_\infty + U_W \right) = \frac{1}{2} U_\infty \left(1+ \sqrt{1-\widetilde{C}_T\cos^2\gamma}\right).
    \label{eqn:uRotor}
\end{equation} 
The wake cross section normal to the streamwise direction is also assumed to be elliptical, with transverse radius $\widetilde{R}_W = R_W \cos{\gamma} $, where $R_{W}$ is the vertical wake radius. Based on the conservation of mass rate in the stream-tube, we obtain
\begin{equation}
  \dot{m} =\rho \pi  \widetilde{R}R U_R =\rho \pi \widetilde{R}_W R_W U_W,
\end{equation}
which gives the final expression for $R_N$ as follows:
\begin{equation}
    R_N ~\dot{=}~ \widetilde{R}_W = R \cos{\gamma}\sqrt{\frac{1+\sqrt{1-\widetilde{C}_T\cos^2 \gamma}}{2\sqrt{1-\widetilde{C}_T\cos^2 \gamma}}}. 
\end{equation}

We then derive the velocity scale $V_N$ and length scale $Y_N$ for the wake deflection due to the  yaw of the turbine. Assuming all the air in front of the rotor including that passing through the rotor-swept area and that flow around the rotor is convected in the transverse direction because of the transverse component of the thrust, the velocity scale $V_N$ for the transverse motion of the wake can be derived as follows: 

\begin{equation}
    V_N = \frac{T_y}{\rho A U_\infty} = \frac{ \frac{1}{2} \rho A \widetilde{C}_T \cos^2 \gamma \sin \gamma U_\infty^2 }{ \rho A U_\infty} =  \frac{1}{2} U_\infty  \widetilde{C}_T \cos^2 \gamma \sin \gamma. 
\end{equation}

The length scale for the wake center location $Y_N$ is 
obtained by multiplying $V_N$ by a typical time scale $D/U_{\infty}$ and shown as follows:
\begin{equation}
Y_N =  D \widetilde{C}_T \cos^2 \gamma \sin \gamma.  
\end{equation}

The velocity scale for the turbine-added turbulence is obtained following \cite{yang_3Dhill_2015}  and \cite{foti2018similarity}, where a velocity scale $U_T$ is defined with the thrust force $T$ and the rotor area $A$ for non-yawed wind turbines, as follows: 
\begin{equation}
    U_T = \sqrt{\frac{T}{\rho A}}. \label{eqn:ut} 
\end{equation}
Substituting equation \eqref{eqn:thrust} into  equation \eqref{eqn:ut} yields, the corresponding $U_T$ for yawed wind turbines is obtained as follows: 
\begin{equation}
    U_T = \sqrt{\rho A \frac{\widetilde{C}_T \cos^2 \gamma U_\infty^2 }{2 \rho A}} = U_\infty \cos{\gamma} \sqrt{\frac{\widetilde{C}_T}{2}}.
\end{equation}

\begin{figure}
	\includegraphics[width=\textwidth]{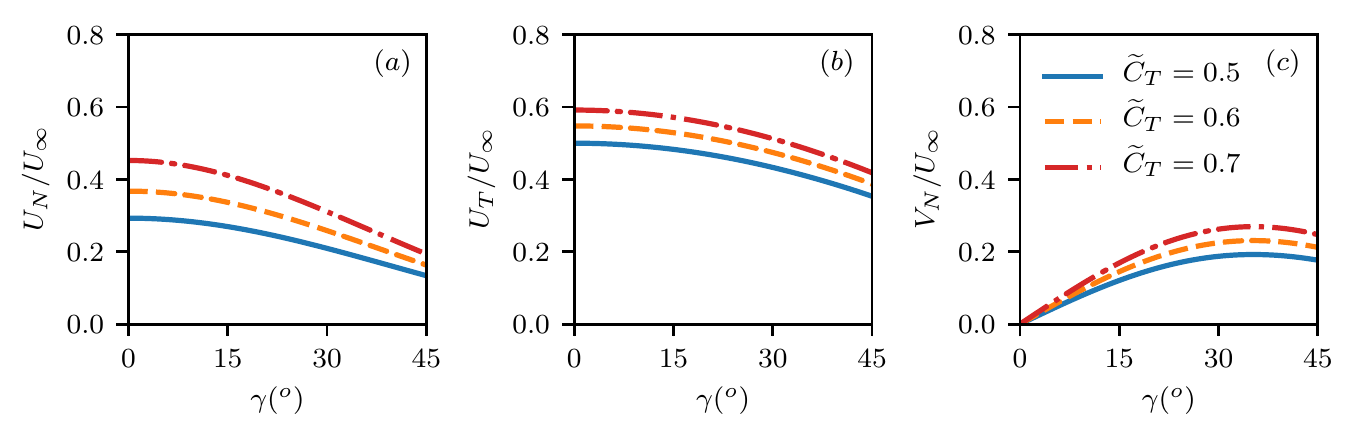}
	\includegraphics[width=\textwidth]{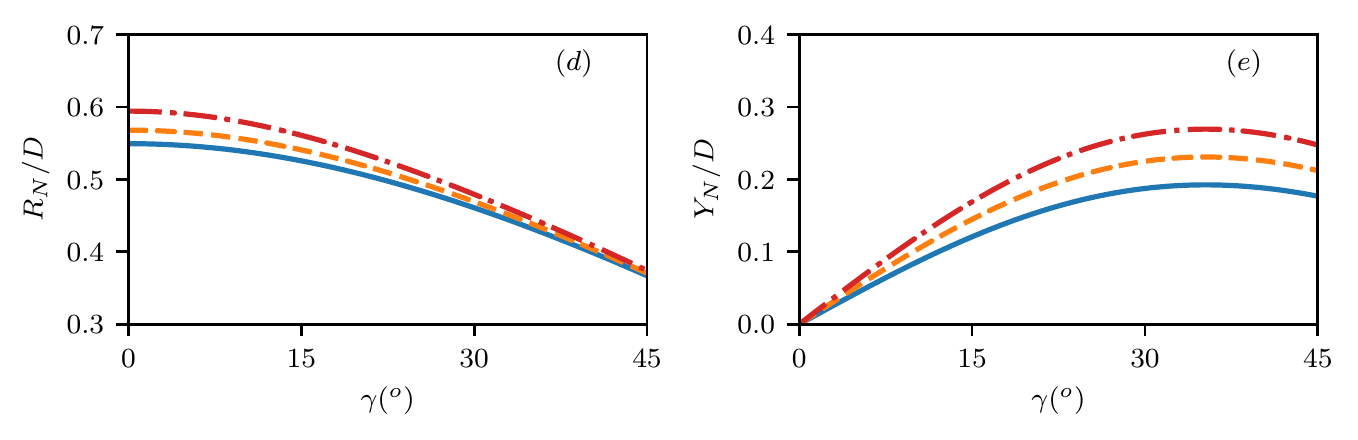}
	\caption{The proposed velocity and length scales varying via yaw angle $\gamma$ for different thrust coefficients $\widetilde{C}_T$.}  
	\label{fig:comparingscales}
\end{figure}

Figure \ref{fig:comparingscales} compares the proposed velocity and length scales, varying via thrust coefficients yaw angle $\gamma$ for different thrust coefficients $\widetilde{C}_T$.  

\section{Wake width in \cite{bastankhah_experimental_2016}}

In equation (6.10) of \cite{bastankhah_experimental_2016}, an analytical expression of wake width is represented with the standard variation of Gaussian curve fit as follows (adapted with the present nomenclature), 

\begin{equation}
    \frac{S}{R} = \cos{\gamma}\sqrt{\frac{1+\sqrt{1-\widetilde{C}_T\cos ^3 \gamma}}{2\left(1+ \sqrt{1-\widetilde{C}_T\cos ^2 \gamma}   \right)}} \approx \cos\gamma \sqrt{\frac{1}{2}} 
    \label{eqn:BPWidth}
\end{equation}
assuming small $\gamma$.

\bibliographystyle{jfm}
\bibliography{biblio}

\end{document}